\documentclass[%
 aip,
 rsi,
 amsmath,amssymb,
 reprint,%
]{revtex4-1}

\usepackage{amsmath,amsfonts,amssymb}
\usepackage{float}
\usepackage{gensymb}
\usepackage{mathrsfs}
\usepackage{subfigure}
\usepackage{url}
\usepackage{natbib}
\usepackage[usenames, dvipsnames]{color}
\usepackage{todonotes}
\usepackage{graphicx}
\usepackage{dcolumn}
\usepackage{bm}

\usepackage[utf8]{inputenc}
\usepackage[T1]{fontenc}
\usepackage{etoolbox}

\makeatletter
\def\@email#1#2{%
 \endgroup
 \patchcmd{\titleblock@produce}
  {\frontmatter@RRAPformat}
  {\frontmatter@RRAPformat{\produce@RRAP{*#1\href{mailto:#2}{#2}}}\frontmatter@RRAPformat}
  {}{}
}%
\makeatother
\begin{document}

\preprint{AIP/123-QED}

\title[Attractor-driven matter]{Attractor-driven matter}
 \email{rahil.valani@adelaide.edu.au}
\author{R. N. Valani}
 \affiliation{School of Mathematical Sciences, University of Adelaide, South Australia 5005, Australia}
\author{D. M. Paganin}%
\affiliation{ School of Physics and Astronomy, Monash University, Victoria 3800, Australia
}%

\date{\today}

\newcommand{\dis}{\ensuremath{\mathcal{D}}}
\newcommand{\raddis}{\ensuremath{\mathcal{R}}}

\begin{abstract}

The state of a classical point-particle system may often be specified by giving the position and momentum for each constituent particle.  For non-pointlike particles, the center-of-mass position may be augmented by an additional coordinate that specifies the internal state of each particle.  The internal state space is typically topologically simple, in the sense that the particle's internal coordinate belongs to a suitable symmetry group.  In this paper we explore the idea of giving internal complexity to the particles, by attributing to each particle an internal state space that is represented by a point on a strange (or otherwise) attracting set. It is of course very well known that strange attractors arise in a variety of nonlinear dynamical systems. However, rather than considering strange attractors as emerging from complex dynamics, we may employ strange attractors to {\em drive} such dynamics.  In particular, by using an attractor (strange or otherwise) to model each particle's internal state space, we present a class of matter coined ``attractor-driven matter''. We outline the general formalism for attractor-driven matter and explore several specific examples, some of which are reminiscent of active matter. Beyond the examples studied in this paper, our formalism for attractor-driven dynamics may be applicable more broadly, to model complex dynamical and emergent behaviors in a variety of contexts.  
\end{abstract}

\maketitle

\begin{quotation}
Strange attractors emerge in the phase space of nonlinear deterministic systems exhibiting chaos. We consider a converse case where attractors, strange or otherwise, are used as a fundamental unit for driving the dynamics of a single-particle or multi-particle classical system.  By coupling the dynamical variables of a particle with an attractor associated with its internal state-space, we present a formalism to generate a class of matter coined ``attractor-driven matter''. We illustrate the rich dynamical and emergent behaviors that can arise from such particles, and show behaviors reminiscent of active matter. The formalism provides a flexible means to generate complex dynamical and collective behaviors that may be broadly applied in various contexts.
\end{quotation}

\section{Introduction}

In classical physics, objects are often modeled as point particles, with the evolution of a single particle or a collection of interacting particles being governed by Newton's second law.  More complicated models endow internal structure to the particles, via internal degrees of freedom such as a magnetic moment or moment-of-inertia tensor. When internal structure is considered to be present, the associated internal state space is typically topologically simple, with examples including (i) a unit-length vector that points in either the up or the down direction (e.g., for Ising-model spin states\cite{SethnaBook}), (ii) a unit vector pointing in any direction in three-dimensional space (e.g., for electric dipole moments\cite{JacksonElectrodynamicsBook}), and (iii) a triad of Euler angles (e.g., for rigid-body orientations\cite{GoldsteinBook1980}). In these three examples, respectively, the state-space coordinate is constrained to lie on (i) the boundary of a 1-sphere, (ii) the boundary of a 3-sphere, and (iii) a 3-torus. Often, but not always, the internal state space of each particle will be constrained by a suitable symmetry group\cite{PismenBook}, with common examples being the set $\mathbb{Z}_p$ of integers modulo $p$ (e.g., for the Potts model\cite{SethnaBook}), the circle group $U(1)$ (e.g., for the phase of a complex scalar optical field\cite{Paganin2006}), and the rotation group $SO(3)$ (e.g., for rigid-body orientations).

While it is not necessary for the ensuing discussion, we restrict consideration to internal state spaces described by a continuum of points.  Often, such an internal state space corresponds to a particular Lie-group symmetry \cite{LipkinBook}, such as the previously mentioned examples of $U(1)$ or $SO(3)$.  What happens when no such symmetry exists, in the internal state space?  To explore this question, take the example of the circle group $U(1)$.  A well-known route to chaos would consider a phase-space limit cycle---which has the same topology as $U(1)$---as bifurcating through a sequence of period-doubling transitions\cite{McCauleyChaosBook}, until a strange attractor\cite{RuelleBook1989B} is obtained.  Inspired by this, suppose the internal state space to be a connected set embedded in $D$ state-space dimensions $\mathbb{R}^D$, which repeatedly bifurcates from the topologically-trivial one-dimensional case of being homotopic\cite{PismenBook} to $U(1)$, into a strange attracting set $\mathcal{A}$ with fractal structure, embedded in $\mathbb{R}^D$.  

\begin{figure*}
\centering
\includegraphics[width=2\columnwidth]{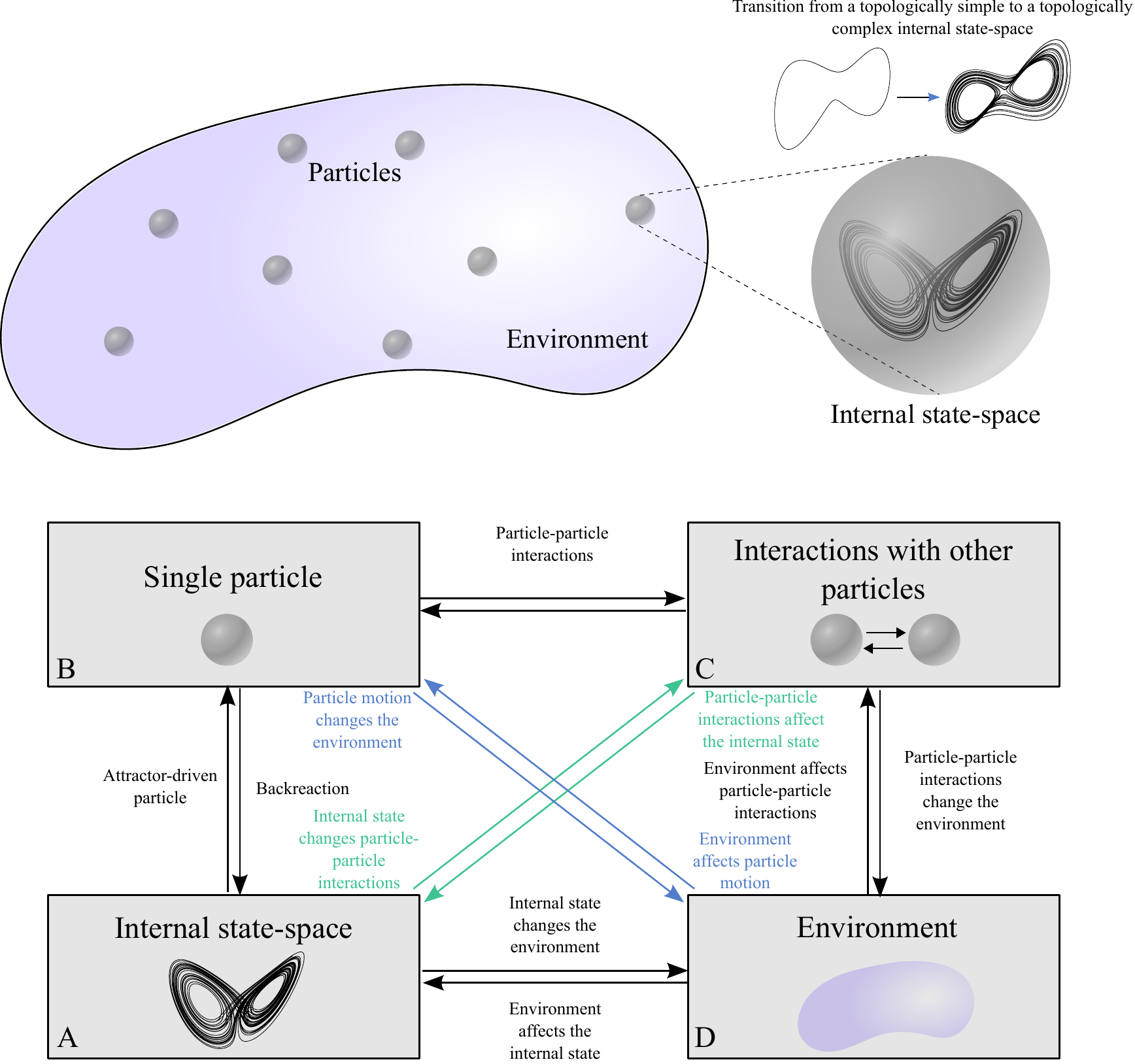}
\caption{Schematic of the formalism for attractor-driven matter. Our system consists of four key elements: (A) the internal state-space of the particle arising from a low-dimensional chaotic dynamical system, (B) single-particle dynamical variables, (C) interactions with other particles, and (D) the environment in which the particles reside. These four elements, when coupled to each other in various ways, can give rise to complex dynamical and emergent behaviors.}
\label{Fig: overview}
\end{figure*}

A rough schematic, of the underpinning idea at the single-particle level, is given in the top right of Fig.~\ref{Fig: overview}.  This sketches a transition from a topologically simple to a topologically complex internal state-space set, in the sense that the latter is a fractal set that may have infinite winding or covering numbers relative to certain points in the space within which the set is embedded.  Actually, as implied by this part of the figure, while we may consider the internal state-space attracting set $\mathcal{A}$ to be fixed in time and space, we may also consider the more general case where the state-space set may itself evolve with time and position.  We are thereby motivated to introduce a control-variable vector  $\boldsymbol{\tau}$, which governs the form of $\mathcal{A}$.  For certain values of $\boldsymbol{\tau}$, $\mathcal{A}$ may be a fractal (strange) set, while for other values of $\boldsymbol{\tau}$, the state-space attracting set $\mathcal{A}$ may be topologically simple (e.g.~homotopic to a circle, with finite winding numbers).   

With the single change that the internal single-particle state space may correspond to a strange attracting set, all of the other classical-mechanics formalism---regarding single-particle and multi-particle systems, the associated interparticle interactions (where present), and the background potential landscape with which the particle or particles may interact---may be carried over without any further alteration.  The logical possibility exists, however, that additional interactions (between individual particles and their environment, or between different particles) may be associated with the strange-attractor internal degrees of freedom.  Figure~\ref{Fig: overview} suggests some of the possibilities that exist in this regard, via the arrows that connect the rectangles in the lower part of the figure.

It is the broader purpose of this paper to promote  exploration of attractor-driven matter, both for single-particle and multi-particle systems. After motivating attractor-driven particles via a physical system in Sec.~\ref{sec: walking droplets}, we outline the general formalism of attractor-driven matter in Sec.~\ref{sec: General formalism}.  As an extended study of a particular special case,  Sec.~\ref{Sec: Active matter} considers attractor-driven matter where the internal state space is considered to be fixed.  Both single-particle and multi-particle instances are examined and connections are made with active matter\cite{PismenActiveMatterBook}. Section~\ref{Sec:EvolvingInternalStateSpace} considers attractor-driven matter for the case where the internal state-space attracting set can evolve, in response to either the environment in which each particle moves (Sec.~\ref{sec: Env}), or the interactions between particles (Sec.~\ref{sec: interactions}). We discuss some broader implications of our paper, together with possible avenues for future work, in Sec.~\ref{Sec: Discussion}. Concluding remarks are given in Sec.~\ref{Sec: DC}.

\section{Motivation for attractor-driven particles}\label{sec: walking droplets}

This section briefly motivates the notion of attractor-driven particles, via the particular example of walking and superwalking droplets.  This specific example gives context to the general scheme, for attractor-driven matter, presented in Sec.~\ref{sec: General formalism}.    

Walking droplets, also known as walkers~\citep{Couder2005WalkingDroplets} (or superwalkers~\citep{superwalker,superwalkernumerical}), are a fascinating hydrodynamical system. In this system, millimetric droplets walk horizontally while bouncing vertically on the surface of a vertically vibrating liquid bath. The walker, upon each bounce, generates a localized slowly decaying standing wave. The walker then interacts with these self-generated waves on subsequent bounces to propel itself horizontally. At high amplitudes of vibrations, the waves created on each bounce decay very slowly in time and the walker's motion is not only influenced by the wave created on its most recent bounce, but also by the waves it created in the distant past, giving rise to memory in the system. In the high-memory regime, walkers mimic several peculiar features that were previously thought to be exclusive to the quantum realm~\citep{Bush_2020}. 

An idealized theoretical model that captures the key dynamics of a walking droplet results in the integro-differential trajectory equation~\citep{phdthesismolacek,Oza2013,Durey2020lorenz,Valaniunsteady2021,Valanilorenz2022}
\begin{align}\label{eq_1}
\kappa\ddot{x}(t)+\dot{x}(t)
={\beta}\int_{-\infty}^{t}\sin(x(t) - x(s))\,\text{e}^{-(t-s)}\,\text{d}s.
\end{align} 
This dimensionless equation of motion describes one-dimensional horizontal dynamics of a walker located at $x$, which continuously generates waves with cosine-function spatial form that decay exponentially in time. The left side comprises an inertial term $\kappa\ddot{x}$ and a drag term $\dot{x}$, with an overdot denoting differentiation with respect to time $t$. The right side quantifies the forcing on the droplet due to the underlying wave field, which is proportional to the gradient of the underlying field. Since this model takes into account the waves generated from all previous impacts, the underlying wave field is calculated through integration of waves generated from all the previous bounces of the walker. The two parameters, $\kappa$ and $\beta$, may be interpreted as the ratio of inertia to drag and the ratio of wave forcing to drag respectively. 

\citet{Valaniunsteady2021} and \citet{Valanilorenz2022} showed that the walker's integro-differential equation of motion in Eq.~(\ref{eq_1}) can be transformed to the following system of ordinary differential equations (ODEs) (see Appendix~\ref{app: walker equation} for a derivation):
\begin{align} \label{Lorenz_droplet}
\dot{x}&=X, \\ \nonumber
\dot{X}&=\sigma\left(Y-X\right), \\ \nonumber
\dot{Y}&=-XZ+rX-Y,\\ \nonumber
\dot{Z}&=XY-bZ. \nonumber
\end{align}
These ODEs are the classic Lorenz equations~\citep{Lorenz1963} coupled with the droplet's motion. The $X$ variable in the Lorenz system is equivalent to the droplet's velocity $\dot{x}$ while the $Y$ and $Z$ variables are related to the memory forcing. The parameters $\sigma, r, b$ in the Lorenz system are related to the walker's system via
\begin{equation}
\nonumber
\sigma=1/\kappa, \quad r=\beta, \quad b=1. 
\end{equation}
Thus, the walker's dynamics given by Eq.~\eqref{Lorenz_droplet} may alternatively be interpreted as an overdamped particle being {\em driven} by the internal state-space Lorenz-system variable $X$. This system forms a motivating example of attractor-driven matter, with a fixed internal state-space governed by the Lorenz system.

\section{Attractor-driven matter: General formalism}\label{sec: General formalism}

Here we outline the general formalism for attractor-driven matter.  Its key elements are indicated by the boxes in the lower part of Fig.~\ref{Fig: overview}.  These four boxes are covered, in turn, in the following four subsections. A fifth subsection considers the mutual interplay, between the aspects of the theory that are described in the preceding subsections.    

\subsection{Internal state-space attractor}\label{sec:GeneralRemarks}

It is common for elementary textbooks on classical mechanics to consider the dynamics of single point particles, subsequently generalizing to ensembles of interacting point particles.\cite{FowlesCassidayBook}  In either case, the state of the system may be specified using suitable generalized coordinates and their corresponding generalized velocities.  Structure can be endowed upon the hitherto-structureless point particles via suitable internal degrees of freedom, e.g.~Euler angles\cite{GoldsteinBook1980} associated with each member of a swarm of rigid bodies.  These additional internal degrees of freedom may be used to augment the system's generalized coordinates and generalized velocities.  If a transition is then made from generalized velocities to canonically conjugate momenta, Hamilton's equations can be employed to study the evolution of the system \cite{GoldsteinBook1980}. 

For a single classical particle, or each member of a system of classical particles, we can augment the  generalized coordinates and velocities (associated with degrees of freedom such as position and mechanical momentum) with a generalized coordinate and velocity corresponding to a state-space vector constrained to evolve along an internal state-space attracting set $\mathcal{A}$.  We speak of $\mathcal{A}$ as a {\em driving} state-space set, because by construction it influences the evolution of its associated particle.  When $\mathcal{A}$ has fractal structure~\cite{McCauleyChaosBook}, we term it a ``driving strange attractor''.  While we are primarily interested in the case where the driving attractor is strange, it is natural to consider the form of $\mathcal{A}$ to depend on the control-parameter vector $\boldsymbol{\tau}$.  Allowing the control parameters to depend on both the position $\mathbf{r}_m$ of the $m$th particle, and time $t$, we may consider $\mathcal{A}_m(\boldsymbol{\tau}_m)$ as classifying each particle $m$ as belonging to one of two equivalence classes\cite{MacdonaldGroupTheoryBook}, corresponding to whether or not $\mathcal{A}_m$ has fractal structure. Other criteria may be employed, for partitioning attractor-driven particles into equivalence classes based on their internal attracting sets.

\begin{figure}
\centering
\includegraphics[width=0.9\columnwidth]{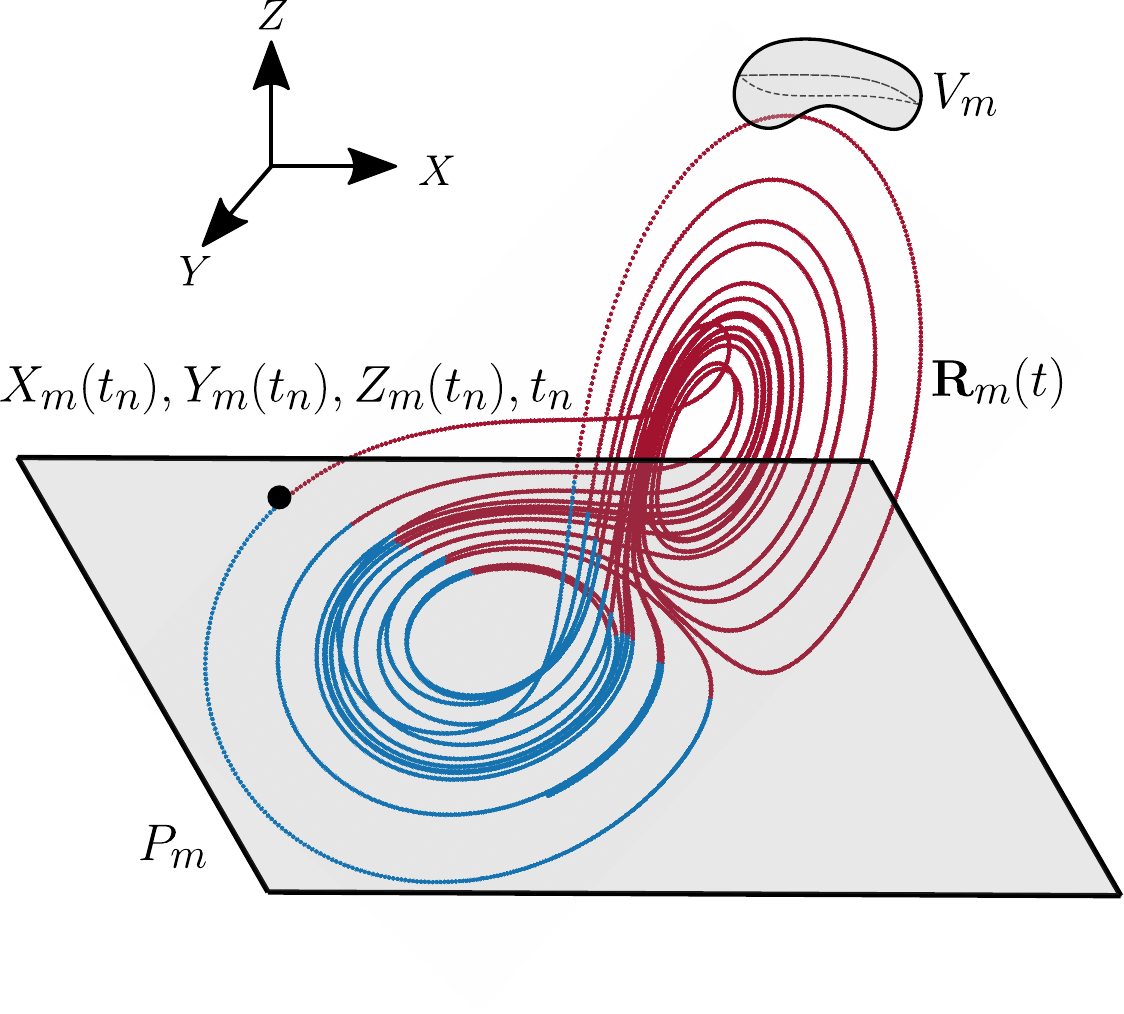}
\caption{Modeling of the internal state-space and particle motion, for the $m$th particle in a system of $N \ge 1$ particles. The internal state-space materializes as an attracting set in the phase-space of a chaotic dynamical system, towards which the state-space trajectory $\mathbf{R}_m(t)$ in Eq.~(\ref{eq:EvolvingStateSpaceVector}) evolves.}
\label{Fig: schemaic}
\end{figure}

Unless stated otherwise, for the remainder of the paper we consider $\mathcal{A}_m$ to be a strange set.  The attractor component of the state space for the $m$th particle is an evolving vector $(X_m(t),Y_m(t),Z_m(t),\cdots)$ which traces out a trajectory in the state space $\mathcal{S}$, along (or asymptoting towards) a driving strange attractor set $\mathcal{A}_m\in \mathcal{S}$. For example, as shown in Fig.~\ref{Fig: schemaic}, consider an otherwise-structureless point particle $m$, possessing a single internal degree of freedom associated with a Lorenz attractor\cite{Lorenz1963,Sparrowbook} having a fixed control-variable vector (cf.~Eq.~(\ref{Lorenz_droplet}))
\begin{equation}
\label{eq:ControlParameterForLorenzAttractor}
(\sigma,r,b)\equiv \boldsymbol{\tau}_m. 
\end{equation}
An associated generalized coordinate, corresponding to the previously mentioned single degree of freedom associated with $\mathcal{A}_m$, is the arc length $s(t)$ that corresponds to the length swept out by the state-space vector\footnote{strictly speaking, in Eq.~(\ref{eq:EvolvingStateSpaceVector}) we should  write `$\sim$' rather than `$\in$'.}
\begin{equation}
\label{eq:EvolvingStateSpaceVector}
\mathbf{R}_m(t)=(X_m(t),Y_m(t),Z_m(t))\in\mathcal{A}(\boldsymbol{\tau}_m)  
\end{equation}
relative to a specified fiducial time. The corresponding generalized velocity is $\dot{s}(t)$, namely the speed at which the attractor set $\mathcal{A}_m$ is traversed. 

\subsection{Single-particle dynamics}\label{sec:SingleParticleDynamics}

There are various means by which the internal state space may influence particle dynamics.  For simplicity, we first consider the case where the internal state-space attractor is independent of time. As shown in Eq.~(\ref{eq:EvolvingStateSpaceVector}), the internal degree of freedom for the $m$th particle is described by the state-space location $(X_m(t),Y_m(t),Z_m(t))$, which moves along its associated internal attractor.  By construction, this generates a continuously-evolving set of numbers that may influence the motion of the particle.  There are many means via which this influence may be modeled, some of which are outlined in the ensuing paragraphs.

At a very general single-particle level, the time-dependent parameters generated by the internal state space include $\mathbf{R}_m(t)$ itself (see~Eq.~(\ref{eq:EvolvingStateSpaceVector})), as well as the time derivatives $\dot{\mathbf{R}}_m(t), \ddot{\mathbf{R}}_m(t), \cdots$.  A Newtonian equation of motion, for the position vector $\mathbf{r}_m(t)$ of the $m$th particle in a system of $N$ particles, is then
\begin{equation}
\label{eq:GenericSingleParticleNewtonLawWithAttractorDriving}
M_m \ddot{\mathbf{r}}_m(t)=F_m[\mathbf{V}_m(t); \mathbf{R}_m(t),\dot{\mathbf{R}}_m(t), \ddot{\mathbf{R}}_m(t),\cdots].
\end{equation}
Here, $M_m$ is the mass of the $m$th particle, whose acceleration is proportional to the function $F_m$ of: (i) the variables $\mathbf{V}_m(t)$ that commonly appear in classical mechanics, such as interparticle interactions (see Sec.~\ref{sec:InteractionsBetweenParticles}) and external fields associated with the environment in which the particle is embedded (see Sec.~\ref{sec:InteractionsWithEnvironment}); (ii) the parameters $(\mathbf{R}_m(t),\dot{\mathbf{R}}_m(t), \ddot{\mathbf{R}}_m(t),\cdots)$ that are generated by the internal state space for the $m$th particle.  At an axiomatic level, $F_m$ can be considered as an arbitrary force function whose form is otherwise unspecified.  For passive physical systems---such as a set of massive point particles whose motion is governed by the gravitational field created by their mutual attraction---specific models can be introduced, to deduce the particular functional form for $F_m$ on the right side of Eq.~(\ref{eq:GenericSingleParticleNewtonLawWithAttractorDriving}), consistent with the particular physical system under consideration.  More freedom in the form of $F_m$ is allowed when extending the class of possible scenarios to include active physical systems, such as active particles \cite{PismenActiveMatterBook} (cf.~Sec.~\ref{sec: active matter}).  For example, at least in principle, for an active particle such as an autonomous robot\cite{Robot1,Robot2,Robot3,Robot4}, one could {\em define} the desired functional form for $F_m$, provided that such a form can subsequently be achieved through suitable constructing and programming of the robot. 

For many systems we may wish to consider, including several of the models that will be treated as specific examples later in the present paper, terms proportional to velocity will be present in the equations of motion. Such terms can be motivated by Stokes-type drag forces~\cite{AchesonFluidDynamicsBook}, as routinely done in the study of active-matter systems.  With this in mind, we may explicitly incorporate a Stokes-type drag-force term $\alpha_m \dot{\mathbf{r}}_m(t)$ into Eq.~(\ref{eq:GenericSingleParticleNewtonLawWithAttractorDriving}), with all remaining forces being absorbed into the force function $G_m$.  Hence  
\begin{align}
\label{eq:GenericSingleParticleNewtonLawWithAttractorDrivingAndDrag}
M_m &\ddot{\mathbf{r}}_m(t)+\alpha_m \dot{\mathbf{r}}_m(t) \\ \nonumber &=G_m[\mathbf{V}_m(t); \mathbf{R}_m(t),\dot{\mathbf{R}}_m(t), \ddot{\mathbf{R}}_m(t),\cdots],
\end{align}
where $\alpha_m$ are real non-negative drag coefficients.  In the small-inertia (overdamped) case where the mass terms $M_m$ may all be neglected and the drag coefficients are all non-zero, $G_m/\alpha_m$ reduces to the velocity of the $m$th particle.  Conversely, if the drag coefficients are all zero but the mass terms are all non-vanishing, and $G_m$ is subsequently identified with $F_m$, Eq.~(\ref{eq:GenericSingleParticleNewtonLawWithAttractorDrivingAndDrag}) reduces to Eq.~(\ref{eq:GenericSingleParticleNewtonLawWithAttractorDriving}).  

In light of the evident freedom that is possible for the choice of force functions $F_m$ or $G_m$, particularly when the class of physical systems is extended to include the possibility of active matter, we now consider an especially simple example. Returning to Fig.~\ref{Fig: schemaic}, we present one mechanism to couple particle motion to its internal state-space attractor. Our method uses Poincar{\'e} surfaces and volumes, in a similar spirit to the idea of suspended flow~\cite{gaspard_1998}. Consider an arbitrary plane $P_m$ that cuts through the state-space attractor for the $m$th particle, together with an arbitrary volume element $V_m$ that has a non-zero measure of the attractor enclosed within it. Let $t_n$ be the times when the trajectory on the attractor intersects the plane $P_m$ for the $n$th time, or enters the volume element $V_m$ for the $n$th time. We may use these times $t_n$ as trigger events for the $m$th particle. Such triggers could then be used to employ a measurable of the attractor at the trigger event to select the corresponding variable of the particle motion after the trigger event.\footnote{Alternatively, or in addition, this attractor-induced trigger may signal the particle to select a corresponding variable based on non-attractor variables such as (i) the state of neighboring particles, and (ii) the potential landscape through which the particle moves. We shall have more to say about such possibilities, later in the paper.} For example, the trigger can be used to change the direction of motion of the particle, where the new direction of motion at time $t_n$ may be a function of (i) the angle between $P_m$ and the tangent to $\mathbf{R}_m(t_n)$, or (ii) the curvature of $\mathbf{R}_m(t_n)$ at $P_m$, or (iii) the time spent within the state-space volume $V_m$, etc. In this way, we can construct a broad class of generic particle motions, driven by an attractor in the internal state-space in an entirely deterministic manner. 

We close this subsection by relaxing the assumption, stated in its opening paragraph, that the state-space attractor does not evolve with time. In particular, variants of our model can be examined, in which the driving strange attractor for each individual particle evolves with time, in a manner that is both deterministic and secular.  For example, if we work with a driving attractor of the Lorenz type, the parameters $\sigma_m(t),r_m(t),b_m(t)$ for the $m$th particle could all be functions of time $t$ that evolve according to a deterministic rule.  This evolution of the driving attractor could be considered secular if its characteristic timescale is long compared the characteristic time $\mathcal{T}$ taken to traverse one ``loop'' of the governing attractor. Moreover, a partitioning into equivalence classes may be induced by the evolution---secular or otherwise---of each individual particle's driving attractor. 

We now give two examples, of the statement in the previous sentence. (i) As a first example, we restate a comment from the previous subsection, that a given population of particles may be divided into two equivalence classes, with a given particle belonging to one class if its driving attractor has fractal structure, and the complementary class otherwise. (ii) For another example, given the set of time-dependent cutting planes $\{P_m(t)\}$---with one such cutting plane for each particle---the ensemble of particles is split into two equivalence classes $\mathcal{C}_1,\mathcal{C}_2$.  These equivalence classes correspond to whether $P_m(t)$ does or does not intersect the secularly-evolving driving attractor $\mathcal{A}_m(t)$ of the $m$th particle.  Stated more precisely, the $m$th particle is in $\mathcal{C}_1$ at time $t$ if 
\begin{equation}
\nonumber
\mathcal{A}_m(t) \cap P_m(t) \ne \emptyset,
\end{equation}
where $\emptyset$ denotes the null set, otherwise the $m$th particle is in $\mathcal{C}_2$.  If there are two or more cutting planes for each particle, the number of induced equivalence classes increases.  

Attractor-driven particles belonging to different equivalence classes may exhibit qualitatively different behavior, with individual particles changing equivalence class when a given cutting plane changes from having a null to a non-null intersection with the driving attractor, or vice versa.  Note, also, that depending on the morphology of the attractor, additional equivalence classes may be induced because the set of all cutting planes that intersect the attractor may itself divide into natural classes.      

\subsection{Interactions between particles}\label{sec:InteractionsBetweenParticles}

We now consider how strange-attractor internal state-space sets may be coupled to one another, in the context of interacting neighboring particles.  
We have already touched on this topic, since, when writing Eq.~(\ref{eq:GenericSingleParticleNewtonLawWithAttractorDriving}), we briefly mentioned that interparticle interactions can be included in the variables $\mathbf{V}_m(t)$.  Let us now partition these variables into (i) the components $\mathbf{W}_m(t)$ (such as an external field and non-attractor-related interparticle interactions) that do not depend on internal-attractor states, together with (ii) the complement of the first partition.  For the $m$th particle in a system of $N$ particles, we may now write Eq.~(\ref{eq:GenericSingleParticleNewtonLawWithAttractorDriving}) as     
\begin{equation}
\label{eq:GenericMultipleParticleNewtonLawWithAttractorDriving}
M_m \ddot{\mathbf{r}}_m(t) = F_m[\mathbf{W}_m(t); \mathbf{Q}_1(t), \mathbf{Q}_2(t), \cdots, \mathbf{Q}_N(t)], 
\end{equation}
where
\begin{equation}
\nonumber \mathbf{Q}_j(t)\equiv 
({\mathbf{R}}_j(t),\dot{\mathbf{R}}_j(t), \ddot{\mathbf{R}}_j(t),\cdots), \quad j=1, \cdots, N.
\end{equation}
Similarly, Eq.~(\ref{eq:GenericSingleParticleNewtonLawWithAttractorDrivingAndDrag}) becomes 
\begin{align}
\label{eq:GenericMultipleParticleNewtonLawWithAttractorDrivingAndDrag}
M_m &\ddot{\mathbf{r}}_m(t)+\alpha_m \dot{\mathbf{r}}_m(t) \\ \nonumber &=G_m[\mathbf{W}_m(t); \mathbf{Q}_1(t), \mathbf{Q}_2(t), \cdots, \mathbf{Q}_N(t)].
\end{align}

Again, a variety of functional forms for $F_m$ and $G_m$ is possible.  We restrict ourselves to two examples, that relate specifically to interactions between particles that directly couple to the driving attractors.  (i) At any time $t$, the respective internal states $\mathbf{R}_a(t) \in \mathcal{A}_a$ and $\mathbf{R}_b(t) \in \mathcal{A}_b$ of particles $a$ and $b$ may be used to generate an interaction potential 
\begin{equation}
\label{eq:InteractionPotentialFirstForm}
V'_{ab}=\epsilon f( d_{ab} ) \, \mathbf{R}_a(t) \cdot \mathbf{R}_b(t). 
\end{equation}
Here, $\epsilon$ is a real coupling constant and $f(d_{ab})$ is a scalar function of the separation $d_{ab} \ge 0$ between particles $a$ and $b$. Typically, $f(d_{ab})$ would tend to a constant when $d_{ab}$ becomes arbitrarily large.  The interparticle force, arising from this potential, gives an additive component on the right side of Eqs.~(\ref{eq:GenericMultipleParticleNewtonLawWithAttractorDriving}) or (\ref{eq:GenericMultipleParticleNewtonLawWithAttractorDrivingAndDrag}). (ii) More generally, at any time $t$, the internal states of particles $a$ and $b$ may be used to form a two-body interaction potential 
\begin{equation}
\label{eq:InteractionPotentialSecondForm}
V''_{ab}=f(d_{ab}) \sum_{\mu}\sum_{\nu}\epsilon_{\mu\nu}  R^{\mu}_a(t) R^{\nu}_b(t). 
\end{equation}
Above, $\epsilon_{\mu \nu}$ is a real rank-two tensor coupling, and the indices $\mu,\nu$ each range over the state-space coordinates $(X,Y,Z,\cdots)$, so that $R_a^1(t)$ is the $X$ component of $\mathbf{R}_a(t)$, $R_a^2(t)$ is the $Y$ component of $\mathbf{R}_a(t)$, etc.  Three-body and higher-order interactions may also be written down, if required.     

\subsection{Interactions with the environment}\label{sec:InteractionsWithEnvironment}

Given the previous section's description of how strange-attractor internal degrees of freedom can be coupled to one another, it is natural to next consider how strange-attractor internal state-space sets may couple to an external environment.  This attractor-based environment will, in general, be superposed with non-attractor-based environments (e.g.~an external gravitational field which couples to the mass of each particle). For simplicity, we focus on interactions with the environment that solely couple to internal-attractor degrees of freedom. 

There are many ways in which the internal state-space sets, of either an individual particle or a system of particles, may be influenced by an external environment. (i) As a first example, let the control parameter $\boldsymbol{\tau}_m$ be a specified time-dependent function of the position $\mathbf{r}_m$ of this particle.  If this control-parameter field has the same functional form for all particles in a system of particles, we may write
\begin{equation}
\label{eq:ControlParameterFieldCalledH}
\boldsymbol{\tau}_m=\mathbf{H}(\mathbf{r}_m,t).
\end{equation}
Here, the otherwise-arbitrary vector field $\mathbf{H}(\mathbf{r}_m,t)$ has the same dimensionality as the control-parameter space for $\boldsymbol{\tau}_m$.  This function $\mathbf{H}(\mathbf{r}_m,t)$ is analogous to an applied field for attractor-driven matter, since its alteration of the driving-attractor form as a function of position and time constitutes an environment, which influences the spatiotemporal evolution of the attractor-driven particle or particles moving through that spacetime environment. Note the indirect nature, of the relation between $\mathbf{H}(\mathbf{r}_m,t)$, and the associated force it induces. (ii) To illustrate the previous general example with a more specific example, adapt Eq.~(\ref{eq:ControlParameterForLorenzAttractor}) to
\begin{equation}
\label{eq:ControlParameterFieldForLorenzDrivingAttractors}
\boldsymbol{\tau}_m
=(\sigma(\mathbf{r}_m,t),r(\mathbf{r}_m,t),b(\mathbf{r}_m,t))
\equiv \mathbf{H}(\mathbf{r}_m,t).
\end{equation}
Thus, for this particular case where the driving attractor arising from the Lorenz system depends on the spatiotemporally-varying control variables $(\sigma,r,b)$, the control-parameter field $(\sigma(\mathbf{r}_m,t),r(\mathbf{r}_m,t),b(\mathbf{r}_m,t))$ plays the role of an external environment coupled to the strange-attractor internal degree of freedom associated with each individual particle. (iii) Suppose each particle $m$ to be a small rigid body, whose orientational state is specified by a triplet of Euler angles $(\Phi^{(1)}_m,\Phi^{(2)}_m,\Phi^{(3)}_m)$, in any one of the numerous available conventions.\cite{GoldsteinBook1980} In this case, Eq.~(\ref{eq:ControlParameterFieldCalledH}) generalizes to the following environment which is sensitive to particle orientation: 
\begin{equation}
\nonumber
\boldsymbol{\tau}_m=\mathbf{H}\big(\mathbf{r}_m,\Phi^{(1)}_m,\Phi^{(2)}_m,\Phi^{(3)}_m,t\big).
\end{equation}
Such an orientation-based control-parameter field might be applied, for example, to generate complex swarming of satellite clusters~\cite{NAG201395}. (iv)  Return consideration to an attractor-driven particle whose internal state space is augmented by a cutting-plane $P$ and its associated induced trigger times $t_n$, as shown in Fig.~\ref{Fig: schemaic}. Suppose the form of the driving attractor to be fixed for all particles at all times.  However, let the cutting plane $P$ be a specified function of position, with this specified function being the same for all members of a set of one or more attractor-driven particles. For the example shown in the figure, where the internal state space has coordinates $(X,Y,Z)$, the cutting-plane field can be specified using 
\begin{equation}
\label{eq:GenericCuttingPlaneField}
a(\mathbf{r},t)X+b(\mathbf{r},t)Y+c(\mathbf{r},t) Z + d(\mathbf{r},t)=0. 
\end{equation}
Here, the real functions $(a(\mathbf{r},t),b(\mathbf{r},t),c(\mathbf{r},t),d(\mathbf{r},t))$ induce a spatiotemporal environment which influences the motion of the attractor-driven particle or particles that move through that environment.  Moreover, if appropriate, the cutting-plane field can be a function of the discrete-particle index $m$.  In this latter case, the cutting-plane field is specified by $(a_m(\mathbf{r},t),b_m(\mathbf{r},t),c_m(\mathbf{r},t),d_m(\mathbf{r},t))$. 

\subsection{Putting it all together}\label{sec:PuttingItAllTogether}

The four preceding subsections have individually focused upon the aspects of attractor-driven matter that are listed in the four rectangles at the bottom of Fig.~\ref{Fig: overview}.  For a general attractor-driven system, all four factors will be operative simultaneously.  Each factor couples to all other factors, as indicated by the twelve arrows joining the four rectangles.  Below, we briefly comment on each of these twelve arrows, using the following abbreviations: $A$\,=\,``internal state space'', $B$\,=\,``single particle'',\footnote{Regarding the distinction between $A$ and $B$, note that $A$ refers to the internal state-space attribute $\mathbf{R}_m$ of a particular attractor-driven particle $m$, with $B$ referring to all of its other single-particle attributes (such as velocity, acceleration, rigid-body orientation, etc.).}  $C$\,=\,``interactions with other particles'', and $D$\,=\,``environment''.  Thus, for example, the label ``$A \rightarrow D$'' examines how aspect $A$ can influence aspect $D$.   

\begin{enumerate}
\item $A \rightarrow B$: This is in many respects the starting point for the notion of attractor-driven matter.  It merely articulates the idea that the motion of each individual particle is influenced by the state of its driving internal attractor.
\item $B \rightarrow A$: The concept of backreaction, familiar from emission-induced recoil in classical mechanics or the Abraham--Lorentz equation in classical electrodynamics\cite{JacksonElectrodynamicsBook}, has evident counterparts for attractor-driven matter.  For example, the state-space control parameter $\boldsymbol{\tau}_m$ may depend on the dynamical state of each particle $m$, in a manner that models the influence of backreaction.  
\item $A \rightarrow C$:  The internal state spaces, of two particles that interact with one another, can influence the form of their associated interaction potential.  For examples, see Eqs.~(\ref{eq:InteractionPotentialFirstForm}) and (\ref{eq:InteractionPotentialSecondForm}).
\item $C \rightarrow A$: The interactions between particles can influence the internal state of each particle.  For example, when two particles collide with one another, they may be damaged (or, more generally, altered) as a result of the collision.  When two attractor-driven particles interact with one another---e.g.~when their spatial separation becomes sufficiently small---each of the particles may have their internal-attractor state space altered by the collision process.  This alteration can occur, for example, by (i) the internal attractors remaining unchanged by the collision, but the location of the internal state-space vectors being altered by the collision, and (ii) the state-space control parameters themselves being altered by the collision.
\item $A \rightarrow D$: The presence of a particle can, via its internal state, influence the form of the environment in which it is embedded. For example, when a bar magnet is placed on a surface whose ``environment'' consists of randomly distributed iron filings, the iron filings will partially align with the field created by the magnet.    
\item $D \rightarrow A$: The environment can influence the internal state space of each attractor-driven particle, as outlined in Sec.~\ref{sec:InteractionsWithEnvironment}.
\item $B \rightarrow C$:  Single-particle motions can result in any specified pair of particles having a sufficiently small spatial separation, such that significant interparticle interactions (``collisions'') become operative. 
\item $C \rightarrow B$: Particle-particle interactions can alter the subsequent motion and locations of each particle. 
\item $B \rightarrow D$:  Single-particle dynamical states may alter the environment through which they move.  Examples include an accelerating electric charge that emits electromagnetic radiation, or an ant leaving a chemical trail as it moves through a landscape.
\item $D \rightarrow B$: By definition, the environment can influence particle motion.  Examples include the harmonic potential associated with simple harmonic motion of a classical point particle, or a small classical test mass that travels through an externally applied gravitational field. 
\item $C \rightarrow D$:  When two electrically-charged particles collide, the resulting acceleration results in electromagnetic radiation that augments the electromagnetic background through which these particles subsequently move.  This example, of how interparticle interactions can influence the environment through which these particles move, may be used to motivate analogs in attractor-driven matter.        
\item $D \rightarrow C$: The environment may influence two-body interactions between particles.  For example, the chemical interaction of two colliding biomolecules moving through a solvent liquid may be influenced by both the chemical composition and temperature of that solvent.  
\end{enumerate}

The four aspects $A,B,C,D$ of attractor-driven matter, when coupled to one another as indicated above, may lead to a variety of emergent behaviors in dynamical systems.  Some specific examples, of the formalism presented here, will be given in Secs.~\ref{Sec: Active matter} and \ref{Sec:EvolvingInternalStateSpace}.

\section{Some examples of attractor-driven matter with fixed internal state-space attractor}\label{Sec: Active matter}

In this section, we consider a particular case of the general formalism in Fig.~\ref{Fig: overview}.  In particular, we explore examples of attractor-driven matter in $1$D and $2$D, where we consider particles driven by a fixed strange-attractor along with various forms of particle-particle interactions. However, before providing quantitative results, we provide some background on active matter and motivate an alternate viewpoint of modeling active-matter-like behaviors, using our attractor-driven framework, in Sec.~\ref{sec: active matter}. Similarities with active matter will also be evident when we present our quantitative results in Secs.~\ref{Sec: AM 1D} and \ref{Sec: AM 2D}, which consider one-dimensional and two-dimensional systems, respectively.

\subsection{An alternate viewpoint for active matter}\label{sec: active matter}

Active matter refers to a large collection of active particles, where each individual particle is a self-propelled entity that consumes energy from its surroundings and converts it into directed motion. Examples of active particles include motile living organisms such as humans, birds, fish or microorganisms such as sperm cells, bacteria and algae, as well as artificial autonomous entities such as active colloidal particles~\citep{PhysRevLett.99.048102}, microrobots~\cite{Palagi2018} and walking droplets~\cite{Couder2005WalkingDroplets,superwalker}. Active matter exhibits emergent collective phenomena such as bird flocks, mammalian herds, fish schools, insect swarms, bacterial colonies, swarming robots~\cite{VICSEK201271,PhysRevLett.75.1226,doi:10.1146/annurev-conmatphys-031119-050752}, motility-induced phase separation (MIPS) in active colloids~\citep{doi:10.1146/annurev-conmatphys-031214-014710} and self-organization of microtubules and motors~\citep{Surrey1167,Ndlec1997}.  
    
Complex locomotion of an individual active particle is typically modeled using a stochastic description. For example, a common minimal active-particle model is run-and-tumble particle (RTP) motion, also known as persistent diffusion~\citep{BALAKRISHNAN1988581} or dichotomous diffusion~\citep{doi:10.1142/S0217979206034881}. Here, in one dimension, the overdamped stochastic dynamics of the active particle is governed by the Langevin equation~\citep{PhysRevE.99.032132} 
\begin{equation*}
        \dot{x}(t)= u\,\sigma_s(t),
    \end{equation*}
where $x$ is the position of the active particle, $u$ is a constant self-propulsion speed and $\sigma_s(t)$ is a dichotomous Markov noise term that flips between $-1$ and $+1$, following a constant-rate Poisson process. In two dimensions, the RTP ``runs'' at constant speed in a fixed direction and then instantaneously ``tumbles'', i.e.~reorients itself by choosing an angle randomly and isotropically~\citep{Hartmann_2020,PhysRevE.101.062120}. Such RTPs undergo ballistic motion over short times, while diffusive motion emerges at long times~\citep{BALAKRISHNAN1988581}. Two other widely used minimal models of active particles are Active Brownian particles (ABPs)~\citep{Romanczuk2012} and Active Ornstein-Uhlenbeck particles (AOUPs)~\citep{PhysRevE.100.022601}.

Although a stochastic description is the norm for modeling the motion of active particles, there are examples of animate and inanimate self-propulsion systems where the motion of the active particle is generated from underlying low-dimensional chaotic dynamics, either explicitly or implicitly. Some of these examples include movement patterns of ants~\citep{cole1991} and mud snails~\citep{Reynolds2016}, locomotion of amoebas~\cite{Miyoshi2001} and worms~\cite{Ahamed2021}, active droplets~\citep{stopgoswim}, autonomous mobile robots~\citep{Robot1,Robot2,Robot3,Robot4} and walking droplets~\citep{Bacot2019,PhysRevE.88.011001}. Moreover, features of active particle motion---similar to random walks and diffusive behavior---can also emerge from deterministic chaotic processes. For example, Brownian-like motion from deterministic dynamics has been studied extensively~\cite{PhysRevA.45.1249,Gaspard1998,BECK1996419,CHEW2002275,MACKEY2006167,Klages2007-ud,1990,Hoover1999,Dorfman1999-sx,castiglione_falcioni_lesne_vulpiani_2008,gaspard_1998}. These models include the motion of a particle subjected to a deterministic but chaotic force. Deterministic diffusive motion also arises in Lorenz-like dynamical systems~\citep{Festa_2002} as well as in systems of delay differential equations~\cite{PhysRevE.84.041105,Albers2022,Muller2022pseudo}. Recently, deterministic diffusion was observed in simulations of walking droplet dynamics governed by an integro-differential equation of motion~\citep{Hubert2019,Durey2020lorenz,Durey2020,durey2021,Valaniunsteady2021,Valani2022ANM}. 

Typically, when one models particle dynamics, the set of accessible states in the internal state space of that particle is topologically simple. For example, in simple conventional models of active matter, the structureless active particle is prescribed a constant self-propulsion velocity. This steady self-propulsion velocity, for the entity being modeled, typically emerges from more complex and typically periodic internal processes that can be attributed to the entity. For example, periodically flapping wings of a bird give rise to its steady flying motion, periodic leg movements in a human give rise to steady walking motion and periodic beating of flagella gives rise to steady swimming for microorganisms. Moreover, the more complex locomotion patterns arising from unsteady movements of active entities are coarse-grained~\cite{Huang_1987,SethnaBook} and modeled using stochastic descriptions in conventional active matter models. For example, the complex trajectories of certain animals during foraging that may arise from complex internal processes is typically modeled using L{\'e}vy flights~\citep{Viswanathan1996}, or the complex interaction of a colloidal particle or a microorganism with the surrounding fluid medium is modeled as Gaussian white noise in ABPs~\citep{lowen2020}. Herein, as we will show in Secs.~\ref{Sec: AM 1D} and \ref{Sec: AM 2D}, by adopting an alternate viewpoint of modeling the internal complexity of an active particle via our attractor in the internal state-space, we obtain dynamics and emergent behaviors reminiscent of active matter.

\subsection{Dynamics and collective behaviors in 1D}\label{Sec: AM 1D}
\subsubsection{Single particle}

\begin{figure}
\centering
\includegraphics[width=\columnwidth]{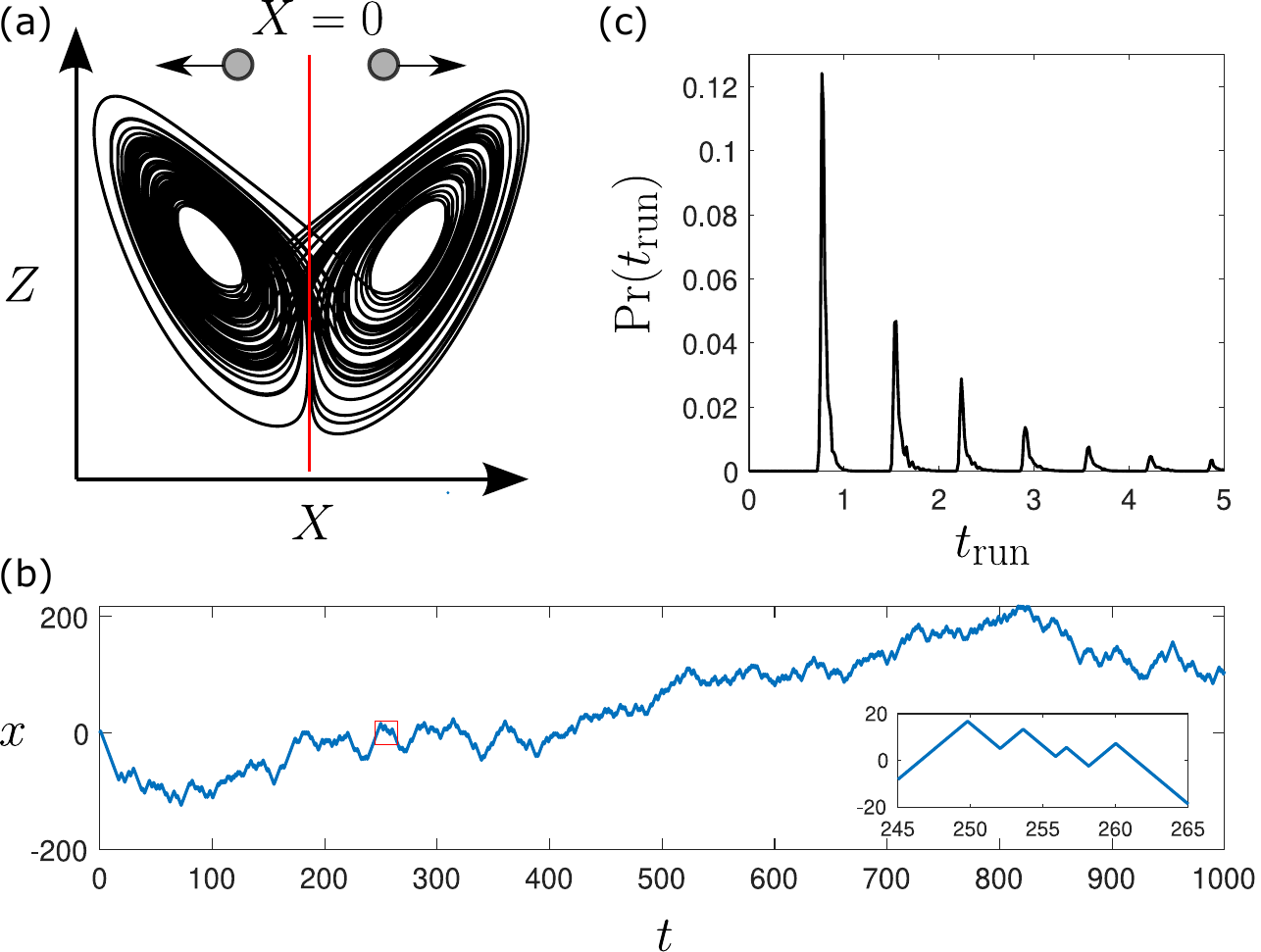}
\caption{Motion of a one-dimensional Lorenz-attractor-driven particle. (a) The Lorenz attractor with parameters $\sigma=10$, $r=28$, $b=8/3$ in the internal state-space drives a particle moving in one-dimension with a constant speed $u=\sqrt{r-1}=3\sqrt{3}$. The particle reverses its direction of motion when the state-space trajectory on the attractor crosses the $X=0$ cutting plane. (b) Typical space-time trajectory. (c) Probability distribution of time spent in constant-speed ballistic motion between direction reversals.}
\label{Fig: 1D RTP Lorenz}
\end{figure}

We start by applying the formalism presented in Sec.~\ref{sec: General formalism}, to explore a one-dimensional attractor-driven particle, which has its internal state-space attractor generated from the Lorenz system. See Appendix~\ref{app: numerical implement} for details regarding the numerical implementation. We coarse-grain \cite{Huang_1987,SethnaBook} the system in Eq.~(\ref{Lorenz_droplet}) by only considering the sign of $X$ for the first equation, giving 
\begin{align}
\dot{x}&=u\,\sigma_d(X), \\ \nonumber
\dot{X}&=\sigma\left(Y-X\right), \\ \nonumber
\dot{Y}&=-XZ+rX-Y,\\ \nonumber
\dot{Z}&=XY-bZ. \nonumber
\end{align}
Here, $\sigma_d(X)=\text{sgn}(X)$ models a fixed cutting plane $P$ located at $X=0$, as described in Sec.~\ref{sec:SingleParticleDynamics}. This gives rise to one-dimensional dynamics where the particle runs to the left or the right with a constant speed $u$ and switches direction when the sign of the variable $X$ flips, i.e.~when a trajectory on the Lorenz strange attractor switches basin of attraction (see Fig.~\ref{Fig: 1D RTP Lorenz}(a)). We see that this motion is reminiscent of an RTP in one-dimension, which has been studied extensively~\cite{PhysRevE.100.052147,Malakar_2018,Singh_2019,Singh_2020,Angelani2014,Angelani_2017}. 

\begin{figure}
\centering
\includegraphics[width=\columnwidth]{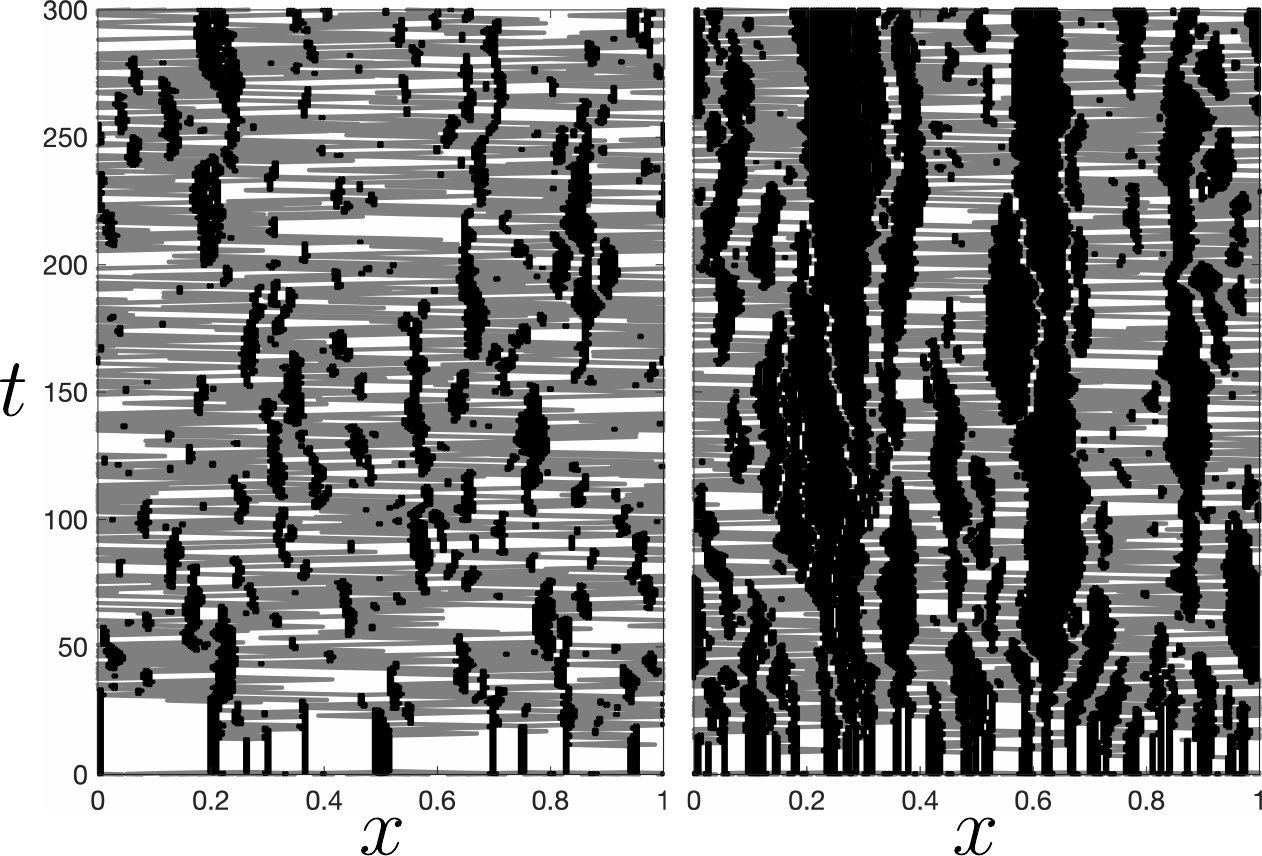}
\caption{Interactions of many Lorenz-attractor-driven 1D particles on the unit interval with periodic boundary conditions. Upon addition of excluded-volume interactions (see Appendix~\ref{app: numerical implement}) in the Lorenz-attractor-driven particles described in Fig.~\ref{Fig: 1D RTP Lorenz} with $u=0.2$, we see the emergence of clustering and jamming. Space-time trajectories show the development of small-scale short-lived clusters for $50$ particles (left), while long-lived large clusters emerge for $100$ particles (right). The gray trajectories denote isolated particles while the black trajectories denote clusters. The size of each particle is $0.005$.}
\label{Fig: 1D many}
\end{figure}

\begin{figure*}
\centering
\includegraphics[width=2\columnwidth]{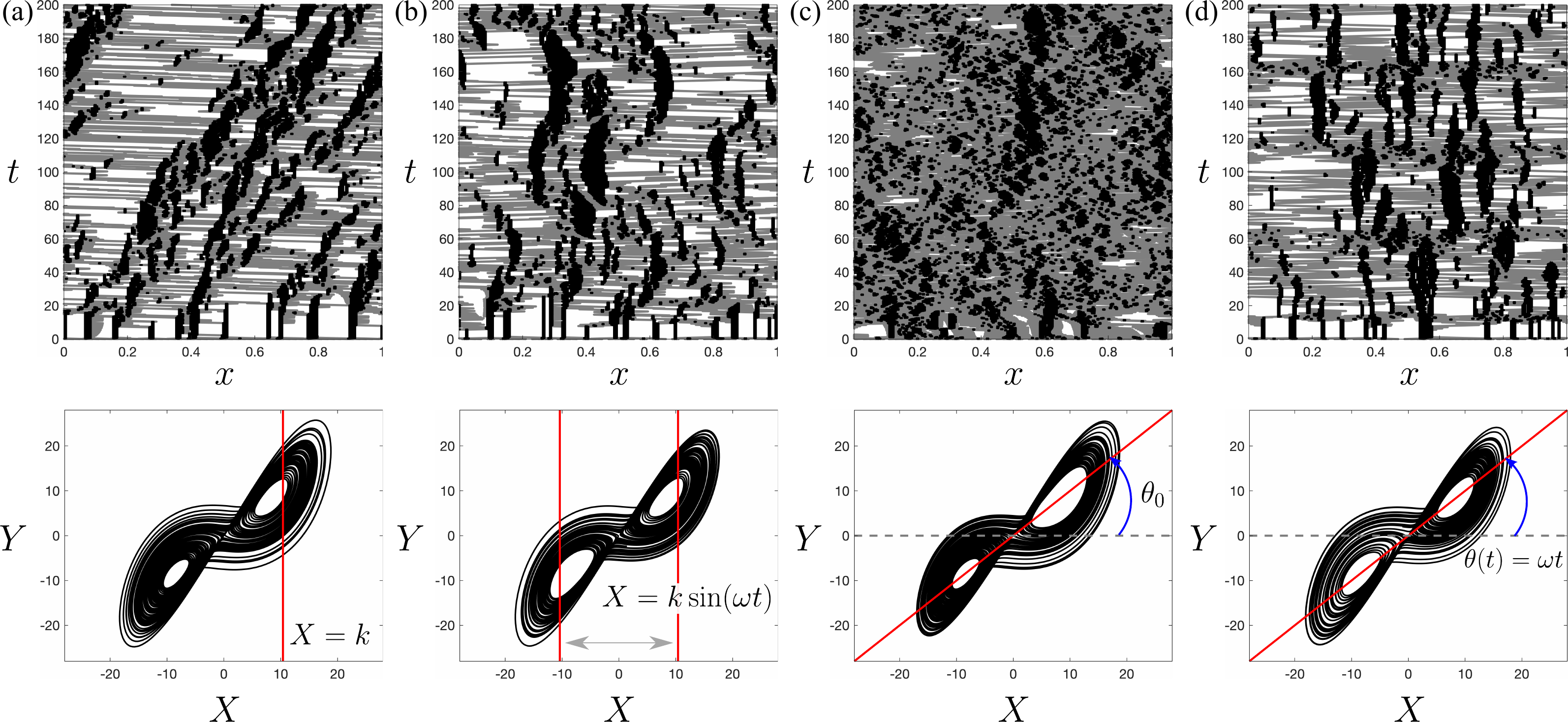}
\caption{Various emergent behaviors (top panels) arising for a collection of $50$ interacting 1D Lorenz-attractor-driven particles (as described in Figs.~\ref{Fig: 1D RTP Lorenz} and \ref{Fig: 1D many}) by keeping the internal state-space attractor fixed and making simple modifications to the cutting plane (red line) that generates particle motion (bottom panels). (a) Offsetting the cutting plane of each particle to $X=k$, where $k=2\sqrt{r-1}$, results in drift of the emergent clusters. (b) Choosing an oscillating cutting plane for each particle of the form $X=k \sin(\omega t)$, with $\omega=2\pi/100$, results in oscillations of the emergent clusters. (c) Changing the cutting plane for each particle to $Y=\tan(\theta_0)X$, with $\theta_0=\pi/4$, gives rise to a different type of clustering with dominance of short-lived small clusters. (d) Allowing the cutting plane to rotate according to $Y=\tan(\omega t)X$, with $\omega=2\pi/100$, results in periodic formation and disintegration of clusters. All other parameters are the same as those employed in Fig.~\ref{Fig: 1D many}.}
\label{Fig: 1D many 2}
\end{figure*}

Figures~\ref{Fig: 1D RTP Lorenz}(b)-(c) show a typical space-time trajectory and the corresponding probability distribution of run durations for this Lorenz-attractor-driven particle. The distribution of run durations is inherent to the internal strange attractor and here results in a distribution with spikes at discrete time intervals, having an exponential decay envelope (see Fig.~\ref{Fig: 1D RTP Lorenz}(c)). These spikes emerge because once the particle switches from one basin of the attractor to another, it performs a discrete number of orbits in that basin, before switching basins again. The characteristic elements of the underlying chaotic flip-flop motion between the two wings of the Lorenz attractor are imprinted on this probability distribution of run durations. For a conventional RTP driven by a Markov dichotomous process, the run durations between direction changes are exponentially distributed~\citep{doi:10.1142/S0217979206034881}. 

Our simple formalism employed with different kinds of attractors in the internal state-space of the particle can generate a wide variety of motions including motion reminiscent of conventional active particles. For example, one can use this formalism to generate ABP-like motion as well as different kinds of active particle motion with anomalous distributions of run durations (see Appendix~\ref{app: 1D active particles}). 

\subsubsection{Many interacting particles}

{We now turn to the question of emergent behavior, in the context of our formalism.} Emergent behavior is a feature of active-matter systems, that arises when interactions are encoded at the level of individual active particles. For example, in one-dimension, interaction of many RTPs with purely repulsive excluded-volume interactions results in clustering and jamming~\citep{PhysRevE.94.022603,PhysRevE.94.022603b,PhysRevE.89.012706,PhysRevE.89.012706}, while introducing simple aligning rules gives rise to one-dimensional flocking states that can stochastically change direction~\cite{O_Loan_1999}. Moreover, interactions induced by a purely repulsive potential can give rise to motile and dynamic clusters for various kinds of active particles in one-dimension~\citep{D0SM00687D}. 

We can readily include such particle-particle interactions that are reminiscent of active matter, in our attractor-driven matter framework. For example, Fig.~\ref{Fig: 1D many} shows spatiotemporal plots of many interacting $1$D Lorenz-attractor-driven particles (described in the caption to Fig.~\ref{Fig: 1D RTP Lorenz}), coupled via an added excluded-volume interaction, with the system being confined on the unit interval with periodic boundary conditions. We see the emergence of clustering and jamming. For smaller numbers of particles we observe short-lived small clusters. Increasing the particle density leads to longer-lived larger clusters. By including short-range spring-like interactions we also obtain dynamic and motile clusters, while introducing aligning interactions results in flocking states that intermittently reverse direction (see Appendix~\ref{app: 1D active particles}). 

\begin{figure*}
\centering
\includegraphics[width=2\columnwidth]{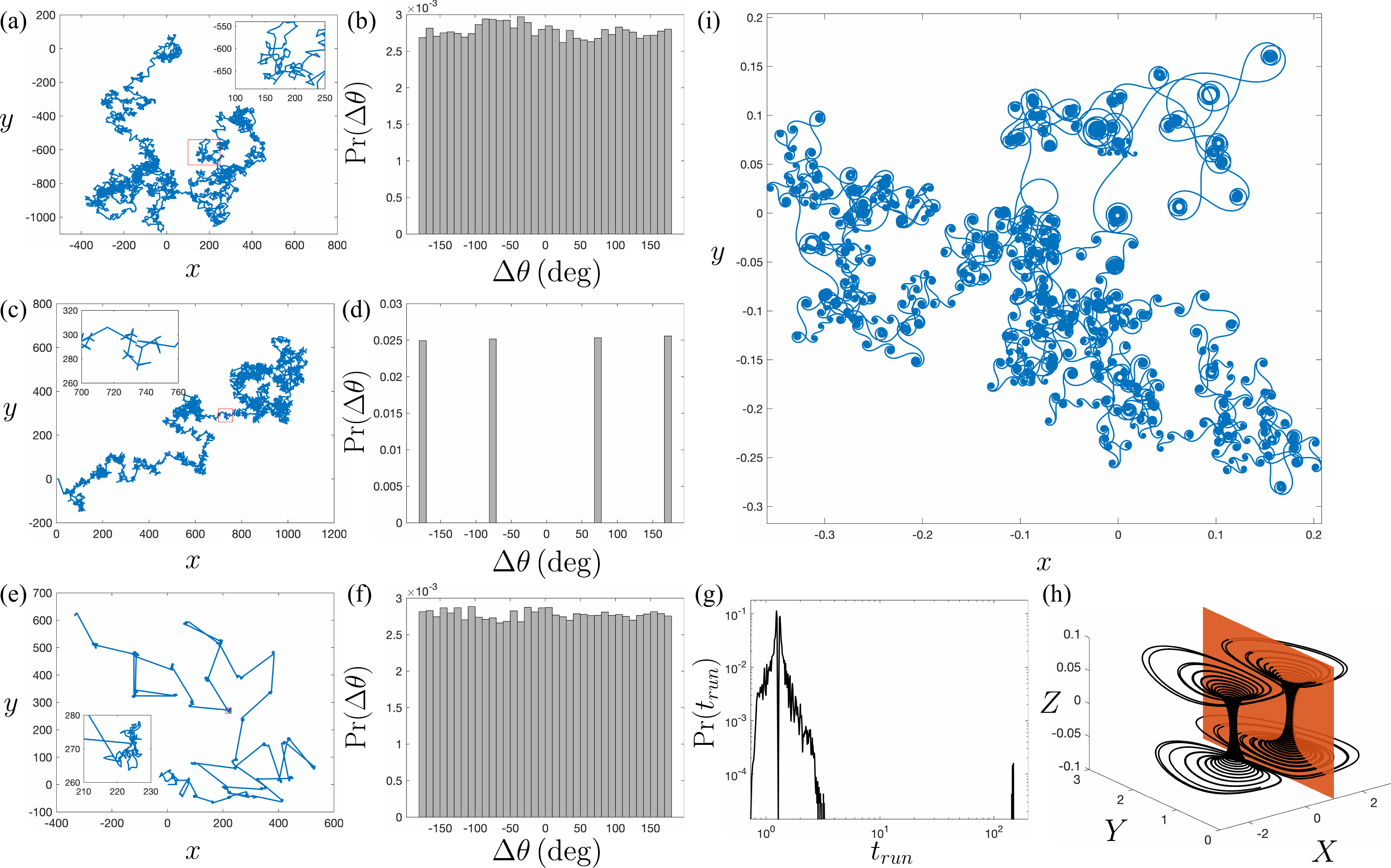}
\caption{Two-dimensional attractor-driven particle motion. (a) A $2$D RTP-like particle trajectory generated from the Lorenz system with parameters $\sigma=10$, $r=28$, $b=8/3$ and self-propulsion speed $u=\sqrt{r-1}$. The cutting plane is $X=0$, which results in a run length distribution that is the same as Fig.~\ref{Fig: 1D RTP Lorenz}(c).  The turning angles are chosen as $\Delta\theta_n=Z(t_n)\,(\text{mod}\,2\pi)$, with the trigger times $t_n$, resulting in a nearly uniform turning angle distribution as shown in panel (b). (c) Trajectory of a particle driven in the same way as in (a) except that (d) turning angles are prescribed to alternate between $\pm\Delta\theta_1=71^{\circ}$ and $\pm\Delta\theta_2=176^{\circ}$ to mimic run-reverse-flick motion. (e) An intermittent particle trajectory generated from the Bouali attractor~\cite{Bouali20143d} ($\dot{X}=3X(1-Y)-2.2Z, \dot{Y}=-(1-X^2)Y, \dot{Z}=0.001X$ with $X=1$ as the cutting plane and self propulsion speed $u=1$) where the particle alternates between a ``long run'' phase of ballistic motion and an ``explore'' phase of diffusionlike motion. For this trajectory, (f) shows the deterministically generated uniform turning-angle distribution, (g) shows the probability distribution of time spent in ballistic motion between direction changes on a logarithmic scale highlighting the two separate time scales, and (h) shows the Bouali attractor along with the cutting plane (red). (i) Trajectory of a particle driven in the same way as in (a) but the turning angles are updated continuously at each instant of time according to $\Delta\theta(t)=Z(t)\,(\text{mod}\,2\pi)$, resulting in an exotic curlicued trajectory with long-time diffusionlike motion.}
\label{Fig: 2D traj}
\end{figure*}

In addition to generating emergent behaviors similar to those realized in conventional active matter, our attractor-driven matter formalism also allows us to readily control and manipulate the emergent behaviors, via relatively simple modifications to the particle motion-generation mechanism. For example, as shown in Fig.~\ref{Fig: 1D many 2}(a), if the cutting plane residing in the internal state-space of each particle is translated to
\begin{equation}
\label{eq:SimpleCuttingPlaneChoice}
X=k>0, 
\end{equation}
then each particle will on average spend more time traveling to the left. Hence, when a cluster is formed, particles are more likely to leave the left side of that cluster and also more likely to join the right side of an adjacent cluster. This results in a net drift of clusters to the right. Thus, by a simple translation of the cutting plane for the Lorenz-attractor-driven particle, one can induce a desired drift in the emergent clusters. Position dependence of this cutting plane, namely a cutting-plane field $P(x)$ described by (cf.~Eq.~(\ref{eq:GenericCuttingPlaneField}))
\begin{equation}
\label{eq:CuttingPlaneField}
a(x)X+b(x)Y+c(x) Z + d(x)=0 
\end{equation}
with $a,b,c,d$ being arbitrary functions of $x$, will enable one to introduce an environment into the system, i.e.~an induced potential landscape in which the clusters' drift direction and speed varies with spatial position $x$. This particular generalization, while interesting, will not be further considered at this point. Rather, we now explore how time dependence in the position of the cutting plane can be used to manipulate the motion of clusters. For example, if the cutting plane is allowed to oscillate between $X=k$ and $X=-k$, then this will induce an oscillation in the cluster dynamics as shown in Fig.~\ref{Fig: 1D many 2}(b). Instead of choosing a vertical cutting plane in the $(X,Y)$ projection of the attractor, if the cutting plane is inclined at an angle $\theta_0=\pi/4$, then the dynamics of emergent clusters and their size distribution changes significantly (see Fig.~\ref{Fig: 1D many 2}(c)). At the cutting-plane angle shown in Fig.~\ref{Fig: 1D many 2}(c), one obtains short lived smaller clusters due to frequent crossing of the inclined cutting plane by the phase-space trajectory. This changing cluster distribution for the cutting plane at different angles can be exploited to model complex clustering behaviors. For example, allowing the cutting plane to rotate in a periodic manner, one can induce periodic formation and disintegration of clusters, as shown in Fig.~\ref{Fig: 1D many 2}(d). 

\begin{figure*}
\centering
\includegraphics[width=2\columnwidth]{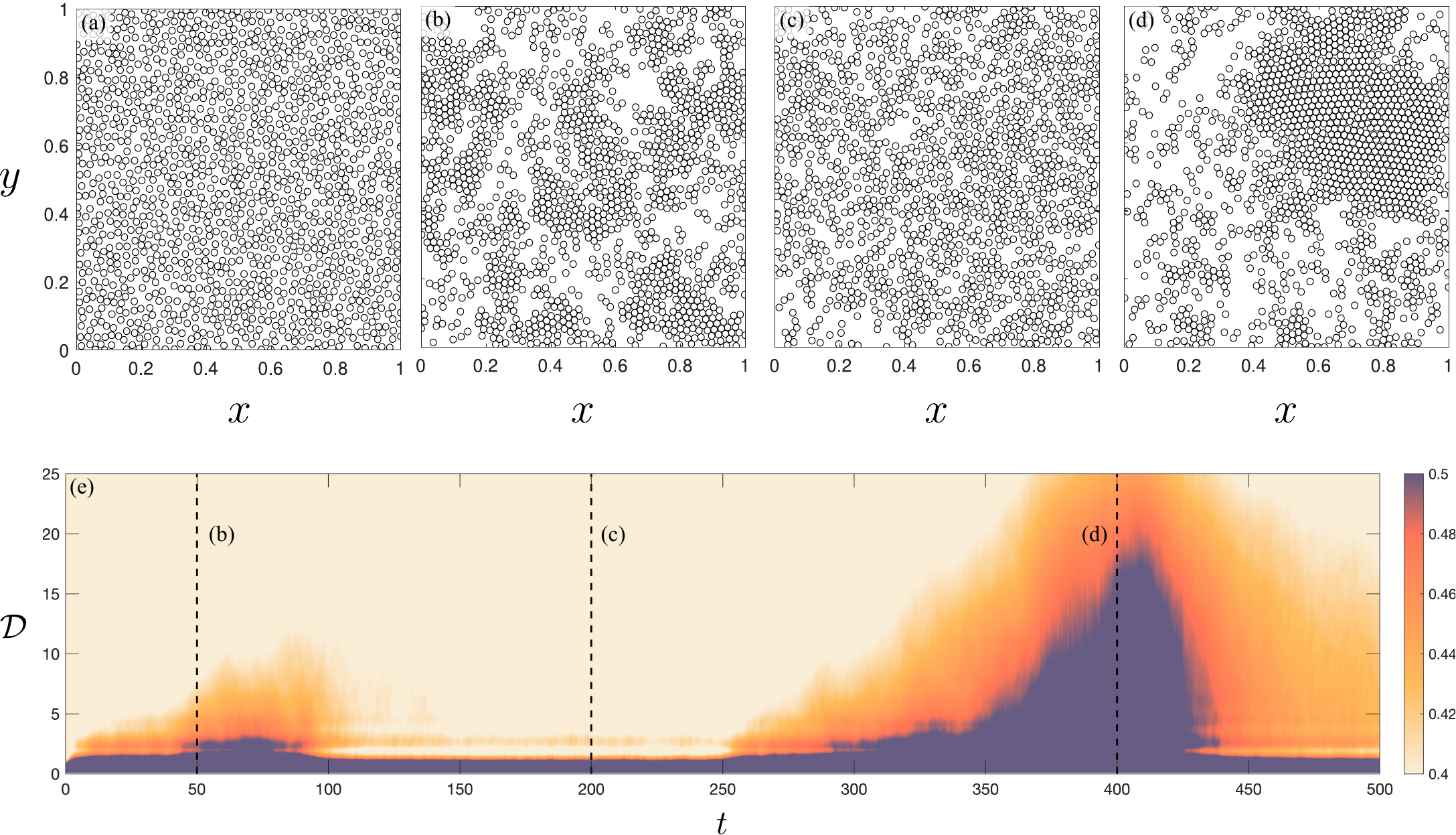}
\caption{(Multimedia view) Emergence of intermittent MIPS-like behavior when $1400$ two-dimensional particles driven by the Bouali attractor (see Fig.~\ref{Fig: 2D traj}(e)-(h)) interact with each other via repulsive harmonic interactions. (a) ($t=0$) Initially randomly located Bouali-attractor-driven particles in the ``run'' phase interact and result in (b) ($t=50$) regions of high and low densities analogous to MIPS. (c) ($t=200$) These  particles then start switching their directions frequently during the ``explore'' phase, which results in the disintegration of clusters. (d) ($t=400$) After some time, the particles again enter the long ``run'' phase and clusters form once more. This cycle keeps repeating as the phase-space trajectory evolves on the Bouali attractor. Panel (e) shows the time evolution of the radial distribution function $\raddis(\dis)$ (filled contours) where $\dis$ is the radial co-ordinate scaled by the particle radius. The parameters for the Bouali attractor are the same as in Fig.~\ref{Fig: 2D traj}(e)-(h). The spring constant for the repulsive forces between the particles is $K=50$ and the radius of each particle is $0.01$. The particles were initiated at random locations in a unit-square periodic domain with slightly different initial conditions ($(X(0),Y(0),Z(0))=(-1,1,0)+\eta(t)$, where $\eta(t)$ is a random number selected uniformly from the interval $[0,0.01]$).}
\label{Fig: 2D Bouali many}
\end{figure*}

We highlight that in all of these examples, the characteristics of the internal state-space---i.e.~the Lorenz attractor, along with the cutting plane---are imprinted in the cluster dynamics and the cluster size distribution.  If the internal-state space attractor is changed, then the corresponding emergent cluster dynamics and statistics will change accordingly.    

\begin{figure*}
\centering
\includegraphics[width=2\columnwidth]{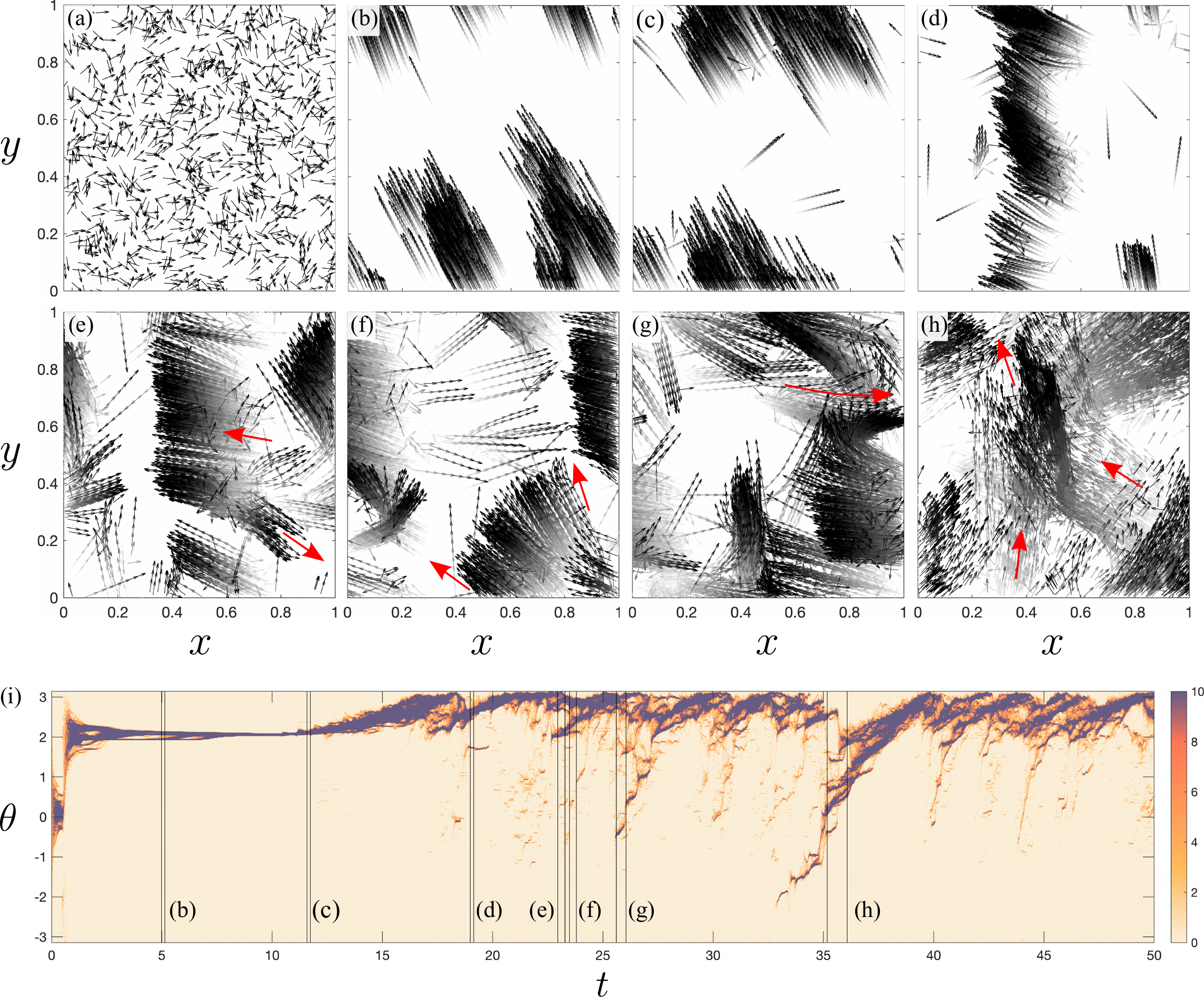}
\caption{(Multimedia view) Complex emergent flocking dynamics in $2$D attractor-driven particles on a periodic unit-square domain. $1000$ Lorenz-attractor-driven particles (as described in the caption to Fig.~\ref{Fig: 2D traj}(a)-(b)) are initialized with random starting positions and directions, along with Vicsek-model-like aligning interactions, as described in the main text. The interaction radius of each particle is $\Delta=0.05$ and the weight factor $W=0.5$. (a) ($t=0$) Initially randomly-located and randomly-oriented particles align and form (b) ($t=5.2$) a coherent flocking phase at early times. At later times we observe more complex flocking dynamics such as (c) ($t=11.8$) a few particles leaving the coherent flock, (d) ($t=19.2$) deformation of the coherent flock into a banded structure, (e) ($t=23.3$) spontaneous ejection of a smaller flock from a bigger flock, (f) ($t=23.8$) flock expansion, (g) ($t=26.1$) scattering of flocks and (h) ($t=36.1$) merging of flocks. The color gradient in the grayscale level shows the past positions of the particles, with the current position indicated in black. The self-propulsion speed of each particle is $u=1$. Panel (i) shows the time evolution of the angle distribution of these particles, with the colorbar representing the number of particles with a given angle. The vertical bars in this panel correspond to the snapshots shown in (a)-(h).}
\label{Fig: 2D many}
\end{figure*}

\subsection{Dynamics and collective behaviors in 2D}\label{Sec: AM 2D}
\subsubsection{Single particle}

We now extend our attractor-driven matter formalism to two spatial dimensions. As an indicative example, in 2D the $x$ and $y$ spatial coordinates of the particle may be taken to evolve via 
 \begin{equation*}
     \dot{\mathbf{x}}=(\dot{x},\dot{y})=(u\,\cos(\theta(t)),u\,\sin(\theta(t))).
 \end{equation*}
Here, $u$ is a constant speed and the angle $\theta(t)$ is now a dynamical variable linked to the internal state-space attractor. We construct 2D attractor-driven particles whose distribution of run durations is dictated by the choice of the cutting plane and the underlying attractor (as in Fig.~\ref{Fig: 1D RTP Lorenz}), and the turning angles $\Delta\theta_n$ are determined by the choice of a measurable of the underlying attractor at the trigger times $t_n$. For a Lorenz-strange-attractor-driven particle with $X=0$ as the cutting plane, we can choose the turning angle in various ways. For example, choosing
\begin{equation}
\nonumber
\Delta\theta_n=Z(t_n)\,\text{mod}(2\pi) 
\end{equation}
results in a nearly uniform distribution of turning angles. A typical trajectory generated in this manner and the corresponding turning-angle distribution are shown in Figs.~\ref{Fig: 2D traj}(a)-(b). This trajectory is reminiscent of $2$D run-and-tumble active particles. 

Instead of determining the turning angles directly from the internal state-space attractor, one can construct a trajectory based on a prescribed probability distribution of turning angles. For example, we can mimic run-reverse-flick motion of an active particle by alternating between two different turning angles. We choose the turning angle to be $\pm\Delta\theta_1$ at trigger times $t_{2k}$ and $\pm\Delta\theta_2$ at trigger times $t_{2k+1}$, where $k$ is any natural number. We simulate two different trajectories in the internal state-space at the same parameter values but different initial conditions. This results in the evolving trajectories  
\begin{align}
\nonumber \boldsymbol{R}^0(t) &= (X^0(t), Y^0(t), Z^0(t)), \\ \nonumber \boldsymbol{R}^1(t) &= (X^1(t),Y^1(t),Z^1(t)). 
\end{align}
When the trajectory $\boldsymbol{R}^0(t)$ hits the cutting plane $X^0=0$, it acts as a trigger to cease the run phase and choose a turning angle according to

\begin{align}
\nonumber
\Delta\theta=\frac{1}{2}\,\text{sgn}(X^1)\Big[(\Delta\theta_2-\Delta\theta_1)\,\text{sgn}(X^0) 
+(\Delta\theta_2+\Delta\theta_1)\Big]. 
\end{align}
By choosing $\Delta\theta_1=71^{\circ}$ and $\Delta\theta_2=176^{\circ}$, we obtain a trajectory similar to run-reverse-flick motion of bacteria~\citep{Taktikos2013}, as shown in Fig.~\ref{Fig: 2D traj}(c)-(d). 

Using different dynamical systems to generate internal state-space attractors, we can generate trajectories of different types. Moreover, instead of determining the turning angles directly from the internal state-space attractors, we can also use the trigger times $t_n$ to act as pseudo-random number generators and generate arbitrary turning-angle distributions (see Appendix~\ref{sec: arbit prob}). For example, Fig.~\ref{Fig: 2D traj}(e) shows an intermittent particle trajectory generated using the Bouali attractor~\cite{Bouali20143d}, with a deterministically generated uniform turning-angle distribution obtained from the trigger times $t_n$ acting as a generator of pseudo-random numbers using the method in Appendix~\ref{sec: arbit prob} (see Fig.~\ref{Fig: 2D traj}(f)). Here the particle alternates between phases of a long ``run'', where it undergoes long-time ballistic motion, and an ``explore'' phase, where the particle performs diffusionlike motion in a localized region of space with relatively short run durations. The probability distribution of the time spent in ballistic motion clearly shows these two separate time-scales (see Fig.~\ref{Fig: 2D traj}(g)), which is an imprint of the crossing statistics of the phase-space trajectory on the Bouali attractor with the specified cutting plane (see Fig.~\ref{Fig: 2D traj}(h)). Another example employs a scaled Lorenz system with a Gaussian turning-angle distribution, to generate a trajectory reminiscent of $2$D ABPs (see Appendix~\ref{sec: 2D ABP}). 

Lastly, instead of choosing the turning angle only at trigger times, one can choose the turning angle continuously from the driving attractor, e.g.~using
\begin{equation}\label{eq: curlicued turning angle}
\Delta\theta(t)=Z(t)\,\text{mod}(2\pi) 
\end{equation}
for the Lorenz strange attractor. This results in a curlicued trajectory with circular and spiral structures, as shown in Fig.~\ref{Fig: 2D traj}(i). 

In this way, one can generate a diverse class of $2$D particle trajectories, including trajectories reminiscent of active particles, whose dynamics and statistics will be characterized by the choice of the underlying attractor, the cutting plane and the turning-angle selection rule. 

\subsubsection{Many interacting particles}

Active particles generally accumulate where they are moving slowly. This results in emergence of motility-induced phase separation (MIPS) where high-density and low-density phases are observed in a large collection of active particles~\cite{doi:10.1146/annurev-conmatphys-031214-014710}. In addition to mimicking conventional MIPS-like phenomena using our attractor-driven framework (see Appendix~\ref{sec: 2D ABP}), we can also generate more exotic MIPS phenomena. For example, interactions of many particles driven by the Bouali attractor~\citep{Bouali20143d} (see Figs.~\ref{Fig: 2D traj}(e)-(h)) result in intermittent MIPS-like phenomena as shown in Fig.~\ref{Fig: 2D Bouali many}. Since the Bouali-attractor-driven particle gives rise to a trajectory where the particle alternates between a long ``run'' phase and an ``explore'' phase, interactions of many such Bouali-attractor-driven particles via harmonic repulsive interactions (as described in Appendix~\ref{app: 1D active particles}) result in cyclical emergent MIPS-like phases. The MIPS-like state is realized when the active particles are in the long ``run'' phase and we observe emergence of clusters. After some time, when the particles enter the ``explore'' phase, their diffusionlike motion results in the disintegration of clusters. This process repeats as particles alternate between the long ``run'' phase and the ``explore'' phase, and we obtain cyclical phases of MIPS-like clustered regions, and gas-like phases with no clear large-scale structure. These MIPS-like clustered regions are quantified by calculating radial distribution functions\cite{SiviaScatteringTheoryBook2011} to reveal particle-particle correlations, as shown in Fig.~\ref{Fig: 2D Bouali many}(e). Here, one can see longer correlation lengths for particles during clustered phases, such as the ones shown in Figs.~\ref{Fig: 2D Bouali many}(b) and (d).

By including aligning interactions between our $2$D attractor-driven particles, we observe the emergence of complex flocking dynamics as shown in  Fig.~\ref{Fig: 2D many}. Here, each particle is driven by the Lorenz attractor residing in its internal state-space, along with the cutting plane $X=0$. To this, we add Vicsek-model-like aligning interactions~\citep{PhysRevLett.75.1226} where each particle aligns itself based on the average direction of the particles in a neighborhood of radius $\Delta$. In our model, each particle gives weight $W$ to the average alignment of its neighbors $\theta_\textrm{nb}$ and weight $1-W$ to its own inherent angle $\theta_\textrm{own}$ arising from the driving strange attractor. Thus, the angle of the $i$th particle at time step $t_k$ evolves according to
\begin{equation*}
    \theta^{i}(t_k)= W \theta^{i}_\textrm{nb}(t_k) + (1-W) \theta^{i}_\textrm{own}(t_k),
\end{equation*}
where 
\begin{equation*}
\theta^{i}_\textrm{nb}(t_k)=\tan^{-1}\left(\frac{\langle \sin(\theta(t_{k-1}) \rangle_{\Delta}}{\langle \cos(\theta(t_{k-1}) \rangle_{\Delta}}\right) 
\end{equation*}
gives the average direction of particles within a circle of radius $\Delta$, and 
\begin{equation*}
\theta^{i}_\textrm{own}(t_k) = \left\{
        \begin{array}{ll}
            Z(t_k) & \quad t_k=t_n \\
            \theta^{i}(t_{k-1}) & \quad \text{otherwise}
        \end{array}
    \right.
\end{equation*}
where $t_n$ corresponds to trigger times and $Z$ is the variable from the Lorenz system (see Appendix~\ref{app: numerical implement} for more details).  Starting with $1000$ particles randomly positioned in a 2D periodic domain with random alignments, we initially obtain coherent motion as shown in Fig~\ref{Fig: 2D many}(b). Once the phase-space trajectory settles on the Lorenz attractor after an initial transient, particles start to leave the coherent flock (see Fig.~\ref{Fig: 2D many}(c)). This results in deformation and change in direction of the coherent flock (see Fig.~\ref{Fig: 2D many}(d)). Further evolution leads to several complex dynamical behaviors such as flock ejection (Fig.~\ref{Fig: 2D many}(e)), flock expansion (Fig.~\ref{Fig: 2D many}(f)), scattering of flocks (Fig.~\ref{Fig: 2D many}(g)) and merging of flocks (Fig.~\ref{Fig: 2D many}(h)). To quantify this complex flocking behavior, we plot the time evolution of the angle distribution of particles as shown in Fig.~\ref{Fig: 2D many}(i). This shows intricate structures that reveal the complex flock interactions described above. 

By incorporating repulsive harmonic interactions in this system along with the Vicsek-like aligning interactions, we get the emergence of $2$D motile crystalline clusters which show features such as void formation and persistence of alignment disturbances within a cluster (see Supplemental Video S1), in addition to the features described in the caption to Fig.~\ref{Fig: 2D many}. 

\begin{figure*}
\centering
\includegraphics[width=2\columnwidth]{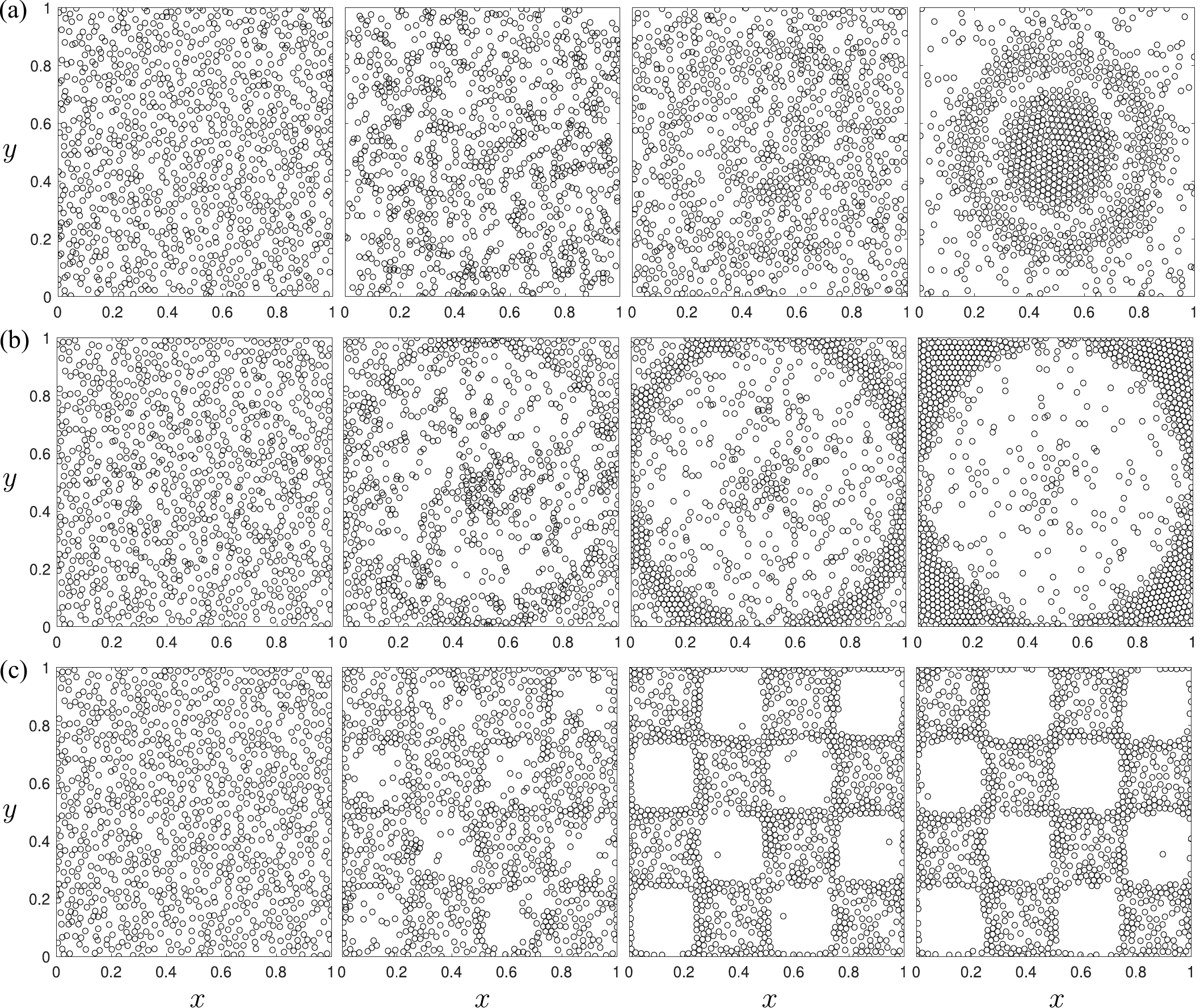}
\caption{Collective dynamics of $1000$ attractor-driven curlicued particles (as presented in Fig.~\ref{Fig: 2D traj}(i)) with repulsive harmonic interactions, where the parameter $r$ of the Lorenz-like system that characterizes the internal state-space attractor varies with the environment, so that $r$ is now the function $r(x,y)$, with (a) (Multimedia view) $r(x,y)=30\sqrt{(x-0.5)^2+(y-0.5)^2}$, (b) (Multimedia view) $r(x,y)=30\,\sin\left(2\pi\sqrt{(x-0.5)^2+(y-0.5)^2}\right)$ and (c) (Multimedia view) $r(x,y)=15\,\sin\left(4\pi(x-0.5)\right) \sin\left(4\pi(y-0.5)\right)$. The snapshots are shown at times (from left to right) $t=0, 0.5, 10$ and $100$. Other parameters are as follows: self-propulsion speed for each particle is $u=0.1\sqrt{r(x,y)-1}$ when $r(x,y)>1$ and zero otherwise, particle radius is $0.01$, spring constant for repulsive interaction is $K=50$ and the internal-state control parameters are $\sigma=5$ and $b=1$. The particle positions were initialized randomly on the unit square with periodic boundary conditions and their internal state-space variables $(X,Y,Z)$ were initialized randomly with values in the interval $[-0.01,0.01]$.}
\label{Fig: Lorenz_env}
\end{figure*}

\section{Some examples of attractor-driven matter with evolving internal state-space attractor}\label{Sec:EvolvingInternalStateSpace}

We now look beyond conventional active matter analogs, to explore some examples of our attractor-driven matter model where the internal complexity---namely the internal state-space attractor---evolves in time. This internal state-space can evolve in various ways. For instance, in our $1$D Lorenz-attractor-driven particle model (shown in Fig.~\ref{Fig: 1D RTP Lorenz}) the internal state-space of the particle can be deformed by gradually evolving the control parameters $\boldsymbol{\tau}=(\sigma,r,b)$ of the driving Lorenz system at each trigger event $t_n$, in a manner governed by the state-space trajectory at each such event (cf.~Eq.~(\ref{eq:ControlParameterFieldForLorenzDrivingAttractors})). This will result in an evolution of the probability distribution of run durations, as the internal state-space attractor evolves.

We explore in detail, in Secs.~\ref{sec: Env} and \ref{sec: interactions} respectively, two specific cases of our attractor-driven matter model, where the internal state-space evolves in time.  In the first of these subsections, the control parameters governing the internal state-space attractor change with the particle position, i.e.~we induce an environment for the internal state-space.  In the second subsection, the control parameters governing the internal state-space attractor change with particle-particle interactions. For both subsections, see Appendix~\ref{app: numerical implement} for details regarding the numerical implementation.

\subsection{Attractor-driven matter with internal state changed by the environment}\label{sec: Env}

In this section, we explore attractor-driven matter where the internal state-space attractor for each particle evolves based on particle position.  This induces an external environment for the internal state-space. Specifically, we consider Lorenz-attractor-driven curlicued particles presented in Fig.~\ref{Fig: 2D traj}(i), with repulsive harmonic interactions between particles, and allow the parameter $r$ of the driving Lorenz system to vary spatially, so that this parameter is now denoted by $r(x,y)$. The phase-space attractor of the Lorenz system correspondingly evolves with changes in the parameter $r$. Moreover, the type of attractor---which may correspond to stable fixed-points, limit cycles or strange attractors---can change when crossing a bifurcation as $r$ evolves. Since the turning angle for these particles is given by Eq.~\eqref{eq: curlicued turning angle} and their self-propulsion speed is $u=0.1\sqrt{r-1}$ when $r>1$ and zero otherwise, for very small $r$ ($r<1$), where $(X,Y,Z)=(0,0,0)$ is a stable fixed-point, we obtain a stationary particle. Hence, spatial regions corresponding to these $r$ values act as attractors for particles. For slightly larger $r$ values ($1<r<15$), the origin becomes unstable and we obtain two stable fixed points at 
\begin{equation}
\label{eq:TwoStableFixedPoints}
(X,Y,Z)=(\pm\sqrt{r-1},\pm\sqrt{r-1},r-1). 
\end{equation}
This typically results in tight circular trajectories since the constant, $Z=r-1$, results in a constant turning angle. We call these ``spinning particles''. However, when $r$ is near values of $2\pi \mathscr{N}+1$ with $\mathscr{N}\in \mathbb{Z}$ in the range $1<r<15$, the turning angle from Eq.~\eqref{eq: curlicued turning angle} is very close to zero and the resulting motion is a straight line or large circular trajectories. Hence, spatial regions corresponding to these values of $r$ act as repellers for particles. For even larger $r$ values ($r>15$) of the internal state-space that correspond to a strange attractor~\footnote{One also occasionally finds small islands of periodicity for certain $r$ values that correspond to a limit-cycle attractor.}, we obtain curlicued trajectories that are qualitatively similar to the one presented in Fig.~\ref{Fig: 2D traj}(i). 

Figure~\ref{Fig: Lorenz_env} shows snapshots of the time evolution of a collection of particles, with three different internal state-space environments. (i) In the first example, as shown in Fig.~\ref{Fig: Lorenz_env}(a), we choose the parameter $r$ to vary radially from the center of the periodic square domain, using  
\begin{equation}
r(x,y)=30\sqrt{(x-0.5)^2+(y-0.5)^2}. 
\end{equation} 
Various effects are at play here, such as the presence of particle-particle interactions, spatially varying $r$ and transient chaos~\cite{Lai2011} that lead to the formation of two clusters: a crystalline solid structure is evident near the center, showing dynamic defects such as lattice vacancies and grain boundaries \cite{KittelBook}, along with a concentric ring-like cluster around it. The gap in the middle is associated with the region of $r(x,y)\approx 2\pi+1$ that acts as a repeller due to the presence of nearly straight-line motion in this region, along with other factors mentioned in the previous sentence.
(ii) In the second example, we choose the parameter $r$ to vary radially, via the sinusoidal form
\begin{equation}
r(x,y)=30 \sin\left(2\pi\sqrt{(x-0.5)^2+(y-0.5)^2}\right).
\end{equation}
As shown in Fig.~\ref{Fig: Lorenz_env}(b), this results in an accumulation of particles near the boundaries of the domain, as well as a few stationary particles near the center of the domain. Here, the formation of a solid crystalline phase near the boundary is correlated with low values of the parameter $r$ (typically $r<1$), since they result in stationary particles. From an alternate viewpoint, one can view this example as a system of particles that self-organizes into a circular domain of gas-like particles contained by a solid-like wall of stationary particles.  (iii) In the third example, $r$ varies according to 
\begin{equation}
r(x,y)=15 \sin\left(4\pi(x-0.5)\right) \sin\left(4\pi(y-0.5)\right). 
\end{equation}
Here, we see an arrangement of particles in a checkerboard pattern, where the particles accumulate in spatial regions corresponding to small $r$ values that support stationary particles (see Fig.~\ref{Fig: Lorenz_env}(c)). However, when the boundaries of these regions become dense and further particles cannot be supported, we find occasional ``wandering'' particles in the spatial regions corresponding to non-stationary particles.

\begin{figure}
\centering
\includegraphics[width=\columnwidth]{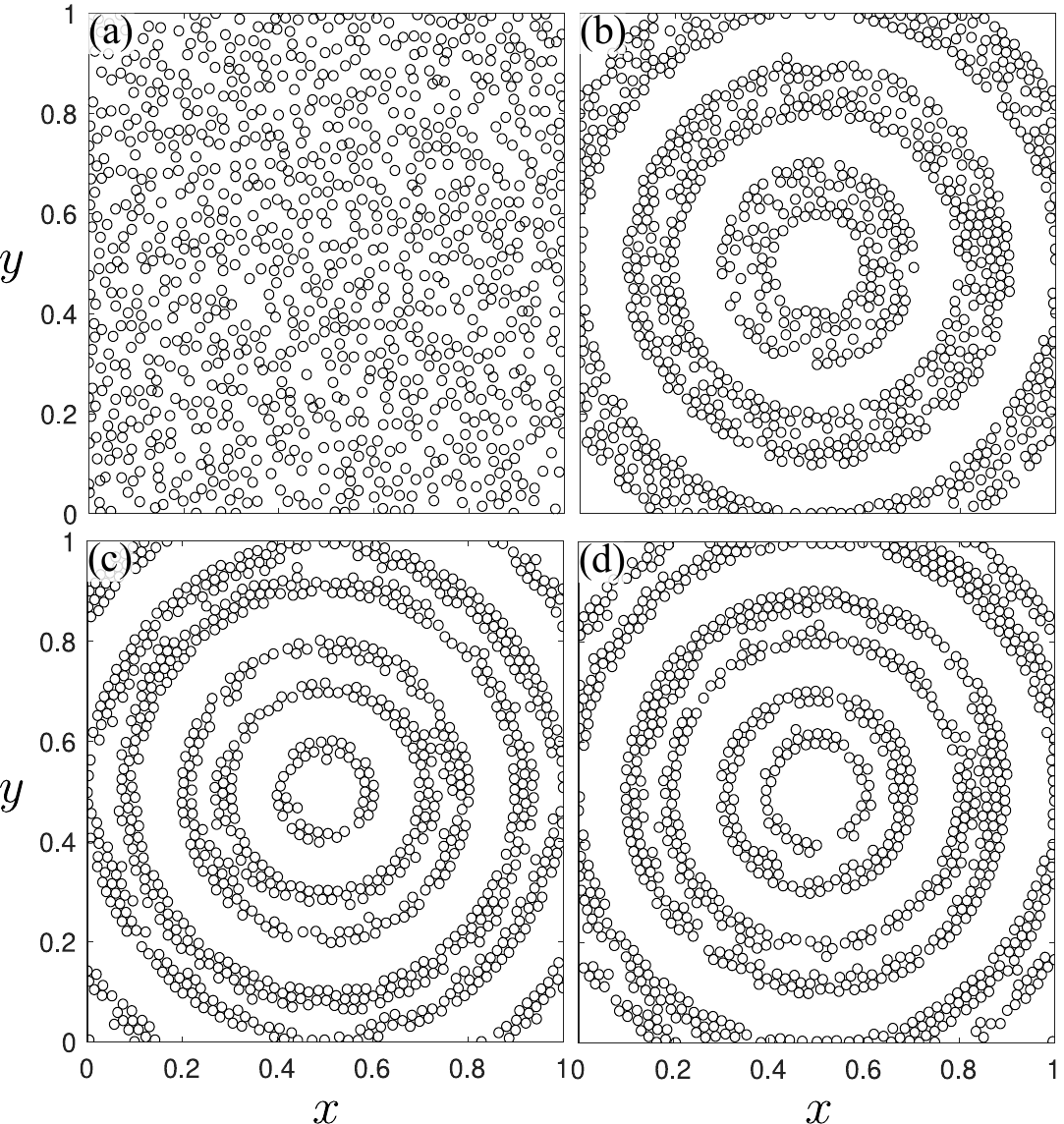}
\caption{(Multimedia view) Emergent behaviors for a collection of $1000$ attractor-driven curlicued particles (as presented in Fig.~\ref{Fig: 2D traj}(i)), with repulsive harmonic interactions, where the internal state-space parameter $r$ of the Lorenz-like system varies with the environment in a time-dependent way, according to $r(x,y,t)=15 \cos\left( \frac{2\pi}{T}t \right)\sin\left(10\pi\sqrt{(x-0.5)^2+(y-0.5)^2}\right)$ with $T=100$. Snapshots are shown at times (a) $t=0$, (b) $t=25$, (c) $t=50$ and (d) $t=100$. Other parameters are as follows: self-propulsion speed for each particle is $u=0.1\sqrt{r(x,y)-1}$ when $r(x,y)>1$ and zero otherwise, particle radius is $0.01$, spring constant for repulsive interaction is $K=50$ and the internal-state parameters are $\sigma=5$ and $b=1$. The particle positions were initialized randomly on the unit square with periodic boundary conditions and their internal state-space variables $(X,Y,Z)$ were initialized randomly with values in the interval $[-0.01,0.01]$.}
\label{Fig: Lorenz_env_time}
\end{figure}

We now consider an example where $r$ varies in both time and space (cf.~Eq.~(\ref{eq:ControlParameterFieldForLorenzDrivingAttractors})). This is shown in Fig.~\ref{Fig: Lorenz_env_time}, for 
\begin{align*}
r(x,y,t) =&15 \cos\left( \frac{2\pi}{T}t \right) \\ &\times\sin\left(10\pi\sqrt{(x-0.5)^2+(y-0.5)^2}\right), 
\end{align*}
with $T=100$. Here, we see that initially randomly distributed particles in the domain accumulate near the regions of small $r$ (see Fig.~\ref{Fig: Lorenz_env_time}(a)-(b)), forming concentric ring-like structures. As time evolves, the $r$ values change at each particle position. Positions corresponding to an increase in $r$ values make the particles mobile and these particles now migrate towards new regions of small $r$, corresponding to stationary states (see Fig.~\ref{Fig: Lorenz_env_time}(c)), with a majority of them accumulating near the boundaries of such regions. Since $r$ varies periodically in time, this results in periodically oscillating concentric ring-like emergent structures.

\begin{figure*}
\centering
\includegraphics[width=2\columnwidth]{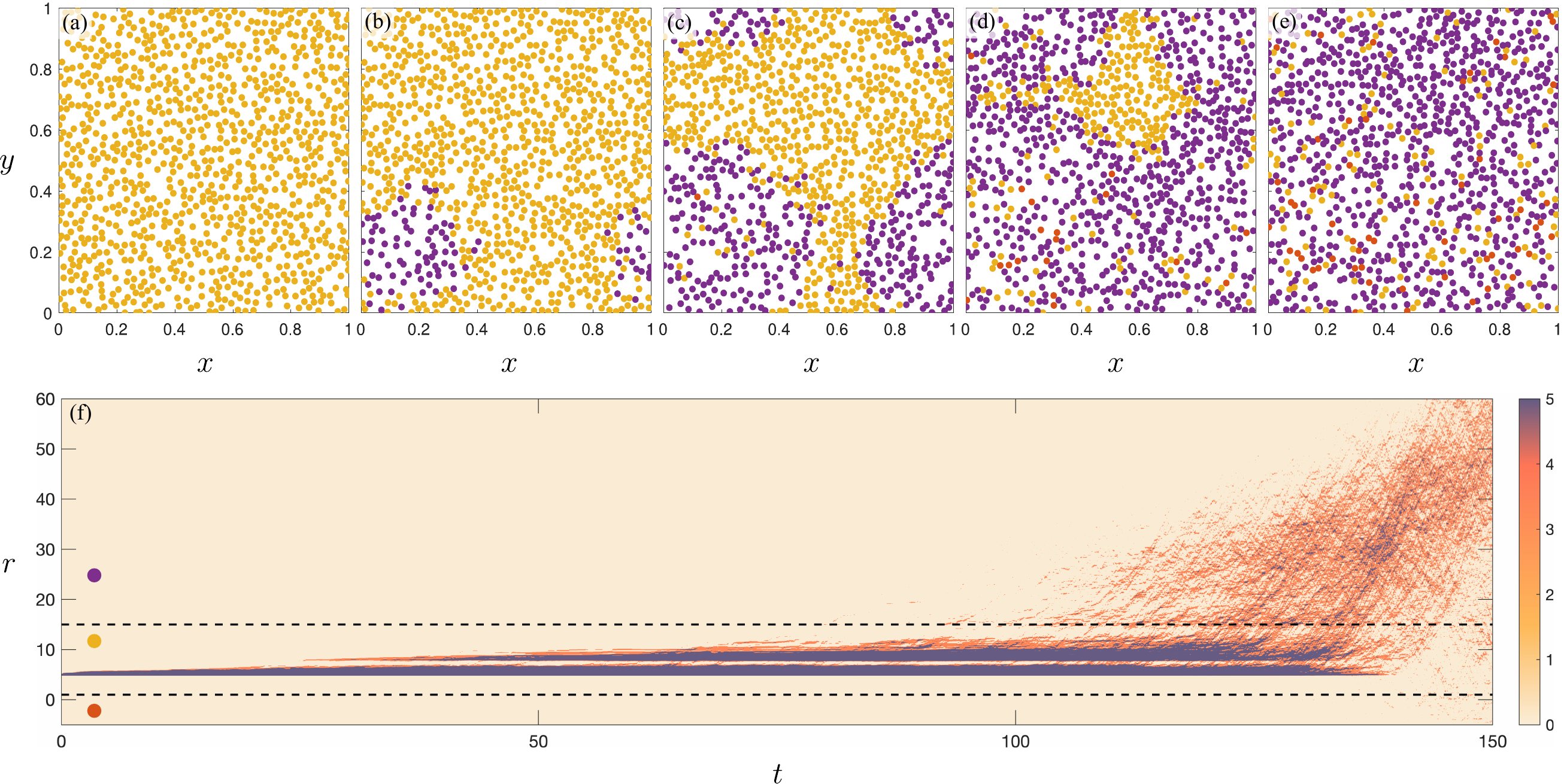}
\caption{(Multimedia view) Emergent behaviors for a collection of $1000$ attractor-driven curlicued particles (as presented in Fig.~\ref{Fig: 2D traj}(i)), with repulsive harmonic interactions, where the internal state-space parameter $r$ of the Lorenz-like system fluctuates due to particle-particle interactions. Snapshots of the particles at times (a) $t=0$, (b) $t=100$, (c) $t=120$, (d) $t=135$ and (e) $t=150$, with the particle color indicating the equivalence class of the internal state-space as follows: red corresponds to stationary particles when the internal state-space has a stable fixed point at $(X,Y,Z)=(0,0,0)$ for $r<1$, yellow corresponds to spinning particles when the internal state-space has a stable fixed point at $(X,Y,Z)=(\pm\sqrt{r-1},\pm\sqrt{r-1},r-1)$ for $1<r<15$, and purple corresponds to curlicued particles where the internal state-space has either an attracting limit cycle or a strange attractor for $r>15$. Panel (f) shows the time evolution of the internal state-space parameter $r$, with the colorbar denoting the number of particles with a given value of $r$. Other parameters are as follows: self-propulsion speed for each particle $i$ is $u_i=0.05\sqrt{r_i-1}$ when $r_i>1$ and zero otherwise, particle radius is $0.01$, spring constant for repulsive interaction is $K=50$ and internal-state perturbation parameter $\epsilon=0.005$. The fixed internal-state parameters are $\sigma=5$ and $b=1$, while the evolving parameter was initialized as $r=5$ for all particles. The particle positions were initialized randomly on the unit square, with periodic boundary conditions, and their internal state-space variables were initialized by adding a small noise $\xi(t)$ to the equilibrium state $(X,Y,Z)=(\pm\sqrt{r-1},\pm\sqrt{r-1},r-1)$, with the values of $\xi(t)$ chosen randomly and uniformly in the interval $[-0.01,0.01]$.}
\label{Fig: Lorenz_interac_twoway}
\end{figure*}

\subsection{Attractor-driven matter with internal state changed by particle-particle interactions}\label{sec: interactions}

Here we consider the case where internal state-space attractors evolve, due to particle-particle interactions. To explore this, we study attractor-driven curlicued particles that interact via repulsive harmonic interactions. When these particles interact, we allow them to perturb each other's $r$ parameter values. For a given pair of particles $i$ and $j$ that are in contact (i.e.~when repulsive harmonic interactions become operative), their respective $r$ values at a given instant of time are updated via
\begin{align*}
    r_i &\rightarrow r_i + \epsilon\,r_j\,\text{sgn}(X_i), \\
    r_j &\rightarrow r_j + \epsilon\,r_i\,\text{sgn}(X_j). \\   
\end{align*}
Here, $\epsilon$ is a small coupling constant governing the perturbation to a particle's internal state. Thus, as particles interact with each other, their respective internal state-space attractors evolve in time.

Figure~\ref{Fig: Lorenz_interac_twoway} shows an example of such a case, where $1000$ interacting particles were initiated randomly in a periodic square domain, all with the same value of the parameter $r=5$. At this $r$ value, the internal state-space attractor has a pair of stable fixed points, located 
according to Eq.~(\ref{eq:TwoStableFixedPoints}), that typically corresponds to a spinning particle (yellow). The local orbiting motion of these particles allows them to occasionally contact their neighbors and thereby perturb their $r$ values. Eventually, this leads to some particles' internal state-space attractors going through bifurcations and becoming strange attractors, resulting in a transition from a spinning motion (yellow) to a curlicued diffusive trajectory (purple)~\footnote{The presence of transient chaos can lead to transient curlicued trajectories for particles whose $r$ values correspond to spinning motion, i.e.~particles indicated by a yellow dot in Fig.~\ref{Fig: Lorenz_interac_twoway}. Moreover, for values of $r$ that are near $2\pi \mathscr{N}+1$ with $\mathscr{N}\in \mathbb{Z}$ in the range $1<r<15$, the turning angle from Eq.~\eqref{eq: curlicued turning angle} is very close to zero and the resulting motion is a straight line or large circular trajectories.}. These curlicued particles contact neighboring particles more frequently and eventually a small cluster of particles emerges, each member of which has a strange-attractor in its internal state-space (see Fig.~\ref{Fig: Lorenz_interac_twoway}(b)). This cluster of curlicued particles---which may also be spoken of as a bubble \cite{SethnaBook} or domain \cite{KittelBook} of one equivalence class\cite{MacdonaldGroupTheoryBook} or phase \cite{SearsSalingerBook} that is surrounded by particles of a different equivalence class or phase---grows in size and eventually fills the entire domain, as shown in Figs.~\ref{Fig: Lorenz_interac_twoway}(c)-(e)). Occasionally, the fluctuations in $r$ values can also lead to some particles becoming stationary if their $r$ values goes below $1$ (red). Note the transition from (i) a smaller entirely-purple domain embedded within a larger entirely-yellow matrix (Fig.~\ref{Fig: Lorenz_interac_twoway}(b)), to (ii) a larger mostly-purple domain (with several yellow ``impurities''\cite{AshcroftMerminBook} and one red ``impurity'') that is embedded in an entirely-yellow matrix (Fig.~\ref{Fig: Lorenz_interac_twoway}(c)), and then (iii) a late stage where all three equivalence classes (purple, yellow, and red) are well mixed, without any obvious domain structure being present (Fig.~\ref{Fig: Lorenz_interac_twoway}(e)).  The corresponding evolution of the $r$-value distribution for the particles, as a function of time, is depicted in Fig.~\ref{Fig: Lorenz_interac_twoway}(f).  The initially-pointlike distribution broadens slightly at early times ($t \lesssim 20$); at intermediate times ($25 \lesssim t \lesssim 70$) the distribution of $r$ values is bimodal, with almost all particles belonging to the ``yellow'' equivalence class; at late times ($t \gtrsim 140$) all three equivalence classes (yellow, purple, and red) are populated, with the majority of particles belonging to the ``purple'' equivalence class.  The system thereby begins with no particles having strange internal attractors, but eventually evolves to a population of particles, where most but not all members have strange driving attractors.

\section{Discussion}\label{Sec: Discussion}

We open this discussion by considering  rare\cite{RareEventsBook1,RareEventsBook2} trigger events, induced by a driving strange attractor $\mathcal{A}$ and an associated cutting plane. This corresponds to an extreme case of Fig.~\ref{Fig: schemaic}, in which (i) only a minuscule fraction $f \lll 1$ of the driving attractor lies on one side of the cutting plane, or (ii) only a minuscule fraction of the driving attractor lies within the trigger volume.  For concreteness, we focus on the case of a cutting plane $\tilde{P}$, with the tilde denoting the fact that this is a ``rare-event cutting plane'', in the sense of the term given in the previous sentence. The induced rare events occur over characteristic timescales that are very long, compared to the timescale $\mathcal{T}$ associated with a segment of the driving attractor that has a winding number on the order of unity (relative to an origin located at the ``center of mass'' of the segment).  For example, for the Lorenz\cite{Lorenz1963,Sparrowbook} or Bouali\cite{Bouali20143d} attractors, $\mathcal{T}$ is on the order of the time taken to traverse one ``loop'' of an attractor lobe.  More generally, $\mathcal{T}$ may be defined as the characteristic time that a state-space trajectory would require, to traverse a state-space arc length $s(t+T)-s(t)$ that is on the order of the diameter $\mathcal{D}$ of $\mathcal{A}$, provided that $\mathcal{D}$ is finite. For example, suppose the driving attractor to be of the Lorenz type, with cutting plane $P\rightarrow\tilde{P}$ in Fig.~\ref{Fig: schemaic} such that only a very small fraction $f \lll 1$ of the strange attractor's measure lies to a particular side of $\tilde{P}$.  This ensures that---for a generic starting point on $\mathcal{A}$---the probability of crossing $\tilde{P}$ is extremely low, over a timescale on the order of the typical time $\mathcal{T}$ taken to traverse one loop of the Lorenz attractor.  Such rare-event triggers could be used to model the finite lifetime of one or more interacting particles, in a variety of settings. For example, in Fig.~\ref{Fig: schemaic}, the cutting plane $P$ could be augmented with the rare-event cutting plane $\tilde{P}$, which models the death (destruction, annihilation, absorption) of a particle.  A similar device could be used to model the birth (creation, emission, binary fission) of a particle.  We should also point out that the cutting-plane or cutting-volume constructions are not necessary to enable rare attractor-induced events.  For example, events with an arbitrarily high degree of rarity could be induced by the condition that the state-space vector lie on a point in the state-space trajectory whose curvature lies between $K$ and $K + \varepsilon$, where $K$ is a finite-sized real number and $\varepsilon$ is a real parameter that can be made arbitrarily small.  

On a different topic, suppose that spatial coarse-graining\cite{Huang_1987,SethnaBook} is employed, so that a large ensemble of motile attractor-driven particles can be viewed as a continuum field \cite{Toner1995}.  In this context, it would be interesting to study if there is any correlation between (i) the different equivalence classes $\mathcal{C}_1,\mathcal{C}_2,\cdots$, as introduced at the end of Sec.~\ref{sec:SingleParticleDynamics}, and (ii) different thermodynamic phases\cite{SearsSalingerBook} of the resulting multicomponent fluid \cite{Bowick2022}. It might also be interesting to investigate the emergent equations of motion associated with the coarse-grained effective continuum field. Even in the absence of spatial coarse graining, the different equivalence classes may behave in a qualitatively different manner, corresponding to different populations of interacting particles, even though the underlying attractor-driven model is by assumption the same for all particles.  We have already seen an example of such behavior, as described in Fig.~\ref{Fig: Lorenz_interac_twoway} and the associated text. 

It would also be interesting to investigate the relation that exists, between the driving-attractor control-parameter field defined in Eq.~(\ref{eq:ControlParameterFieldCalledH}), and the associated force that this control-parameter field induces.  As pointed out in the paragraph containing Eq.~(\ref{eq:ControlParameterFieldCalledH}), such a relation is indirect.  This is because the control-parameter field constrains the evolution of the internal-attractor degree of freedom, which in turn determines the force that is experienced by the particle, as a function of both its internal state and the local environment in which it is immersed. Further, if one estimates the characteristic internal timescale $\mathcal{T}$ discussed in the opening paragraph of this section, the induced force on the attractor-driven particle could be temporally coarse-grained over times that are significantly larger than $\mathcal{T}$.   

We also recall that a number of coherent structures were observed to emerge in our simulations. This invites comparison with the emergent structures observed in conventional active-matter models in which individual particles are driven by stochastic dynamics. In addition to understanding the similarities in the emergent structures arising from both formalisms, it would be interesting to explore any differences, particularly when those differences can be attributed to the attractor nature of the driving.  Similar remarks apply to the possibility for phase transitions in multiple-particle systems of attractor-driven particles, whether they be active or passive in nature. 

While the language of physically-realizable attractor-driven systems has been used throughout this paper, its formalism may also be applicable to virtual systems such as proliferating computer viruses on highly-connected graphs\cite{ComputerVirus2005,ComputerVirus2018,ComputerVirus2019} or in modeling of epidemics on networks~\cite{Kiss2017}.  Our core idea might also be applied to discrete dynamical systems, whether they be physical or virtual.  Here, the internal state space might be given, for example, by the output of a logistic map.

Lastly, one can exploit additional properties of chaotic dynamical systems, to broaden the framework presented in this paper. We provide a few examples here: 
\begin{itemize}
    \item Multistability~\cite{Kapitaniak2015,2021} can emerge in certain dynamical systems where multiple attractors coexist, each with its own basin of attraction. If such a system is employed to generate internal state-space attractors, then depending upon the initial conditions, qualitatively different types of internal state-space attractors will be realized and they will give rise to different equivalence classes of particles. Moreover, sometimes an infinite number of attractors can coexist in a dynamical system, with the notion of having a countably infinite number of attractors being termed  ``megastability''~\cite{Sprott2017} and the notion of an uncountably infinite number of attractors being known as ``extreme multistability''~\citep{Pisarchik2022}. 

    \item Typically, strange attractors are associated with chaotic dynamics, but strange non-chaotic attractors can also exist~\cite{GREBOGI1984261}. Such attractors typically arise in quasiperiodically forced systems~\cite{Feudel2006}. Having such strange non-chaotic attractors in the internal state space will allow us to preserve the fractal nature of the attractor, while discarding sensitivity to initial conditions.

   \item A new type of chaotic behavior termed ``laminar chaos'' was identified in dynamical systems with time-varying delay~\citep{Lamchaos2018}. Such chaotic behavior is characterized by constant-intensity laminar phases that are periodically interrupted by irregular bursts. The intensity levels of the laminar phases vary chaotically but are governed by a one-dimensional iterated map defined by the nonlinearity of the delayed feedback. Features of laminar chaos are robust, being preserved even when noise is added to the system~\cite{Muller2020,Muller2022pseudo}. Using systems that exhibit laminar chaos, to model our internal state-space, will allow us to construct attractor-driven particles that preserve correlations even when disturbed by external noise.

   \item In normal chaos, small changes in initial conditions lead to very different trajectories but the statistics normally remain the same. Recently, a new concept named ultra-chaos~\citep{ultrachaos1} has been proposed, where small disturbances can lead to large deviations in the statistics. Ultra-chaos was reported in the motion of a mobile robot~\citep{ultrachaos2}. Our attractor-driven-matter framework could be readily applied in this context, where an appropriately chosen internal-state-space attractor can generate the required properties (e.g.~ultra-chaos or normal chaos) in the motion of the mobile robot, tailored to the specific application. Moreover, the facet of an evolving internal-state-space attractor in our framework can add an extra layer of complexity, in the chaotic motion of the mobile robot.    
   
\end{itemize}

\section{Conclusion}\label{Sec: DC}

We presented a simple formalism for particles that are driven by internal-state-space attractors, strange or otherwise, and showed examples of some of the many phenomena it can model. The motions and emergent collective behaviors generated using this framework have imprinted in them signatures of the underlying attractors, which constitute the internal state space of each particle in our model. Our formalism might be applied more broadly, beyond the examples given in the present paper.  

\section*{Supplementary Material}

See the supplementary material for videos and $\mathtt{MATLAB}$ scripts related to the simulation results presented in this manuscript.
 
\begin{acknowledgments}
We are grateful to the anonymous reviewers for their critical feedback, which led us to improve our manuscript. R.V.~was supported by Australian Research Council (ARC) Discovery Project DP200100834 during the course of the work.
\end{acknowledgments}

\section*{Conflict of Interest}
The authors have no conflicts to disclose.

\section*{Data Availability Statement}
The data that support the findings of this study are available
from the corresponding author upon reasonable request.

\appendix

\section{Transforming the walker's integro-differential equation into a Lorenz-like system of ODEs}\label{app: walker equation}

Here we provide a derivation to transform the walker's integro-differential equation of motion (Eq.~(\ref{eq_1}))  into a system of ODEs that take the form of the Lorenz system, as originally presented in~\citet{Valaniunsteady2021} and~\citet{Valanilorenz2022}. Consider the walker's integro-differential equation of motion,
\begin{equation*}\label{App: eq_1}
\kappa\ddot{x}(t)+\dot{x}(t)
={\beta}\int_{-\infty}^{t}\sin(x(t) - x(s))\,\text{e}^{-(t-s)}\,\text{d}s.
\end{equation*}
We denote the wave-memory force by
\begin{equation*}
Y(t)={\beta}\int_{-\infty}^{t}\sin(x(t) - x(s))\,\text{e}^{-(t-s)}\,\text{d}s.
\end{equation*}
Differentiating $Y(t)$ with respect to time and using the Leibniz integration rule, we have
\begin{align*}
\dot{Y}(t)&=-{\beta}\int_{-\infty}^{t}\sin(x(t) - x(s))\,\text{e}^{-(t-s)}\,\text{d}s\\ \nonumber
&+{\beta \dot{x}}\int_{-\infty}^{t}\cos(x(t) - x(s))\,\text{e}^{-(t-s)}\,\text{d}s\\ \nonumber
&=-Y(t)+X W(t),
\end{align*}
where $\dot{x}=X$ and 
\begin{equation*}
W(t)={\beta}\int_{-\infty}^{t}\cos(x(t) - x(s))\,\text{e}^{-(t-s)}\,\text{d}s.        
\end{equation*}
Differentiating $W(t)$ with respect to time, we get
\begin{align*}
\dot{W}(t)&={\beta}-{\beta}\int_{-\infty}^{t}\cos(x(t) - x(s))\,\text{e}^{-(t-s)}\,\text{d}s\\ \nonumber
&- {\beta\dot{x}}\int_{-\infty}^{t}\sin(x(t) - x(s))\,\text{e}^{-(t-s)}\,\text{d}s\\ \nonumber
&={\beta}-W(t)-X Y(t).
\end{align*}
By making a change of variables
\begin{equation*}
Z(t)={\beta}-W(t), 
\end{equation*}
we obtain the following system of Lorenz-like ODEs for the walker's dynamics:
\begin{align*}
    \dot{x}&=X\\ \nonumber
    \dot{X}&=\frac{1}{\kappa}(Y-X)\\ \nonumber
    \dot{Y}&=-Y+{\beta}X-XZ\\ \nonumber
    \dot{Z}&=-Z+XY.\\ \nonumber
\end{align*}

\section{Details of numerical implementation}\label{app: numerical implement}

The simulation results presented in Sec.~\ref{Sec: Active matter} of this paper were performed by first solving the chaotic dynamical system for the internal state space, using the inbuilt ode45 solver in $\mathtt{MATLAB}$, and then coupling it with the particle's motion using an explicit Euler method. Alternatively, the simulation results presented in Sec.~\ref{Sec:EvolvingInternalStateSpace} were performed by simultaneously solving the chaotic dynamical system for the internal state space and its coupling with particle dynamics. Here, a fourth order Runge-Kutta method was used for the former, while an explicit Euler method was used for the latter.  We have provided example $\mathtt{MATLAB}$ codes in the Supplemental Material, that simulate some of the behaviors presented in this manuscript. In this section, we provide more details for the implementation of the interactions between attractor-driven particles.

 \begin{figure}
\centering
\includegraphics[width=\columnwidth]{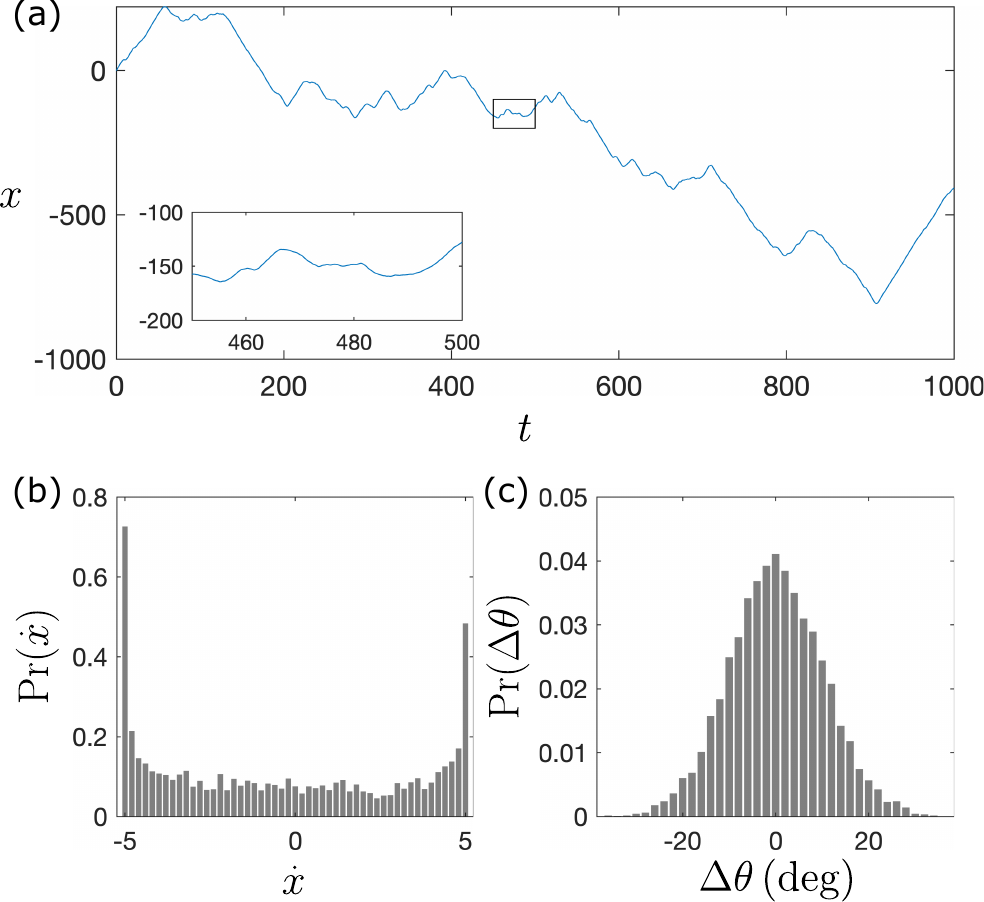}
\caption{Deterministic one-dimensional ABP-like particle generated from a scaled Lorenz system with parameters $\sigma=10$, $r=28$, $b=8/3$ and $F=10$. The one-dimensional trajectory of the particle is shown in (a), while (b) and (c) show the distribution of the particle velocity and the change-in-angle $\Delta\theta$ respectively. Here the constant speed $u=\sqrt{r-1}$.  The mean of the Gaussian distribution for $\Delta\theta$ was chosen to be zero, while the standard deviation was chosen to be $\sigma_{\theta}=10^{\circ}$. 
}
\label{Fig: ABP}
\end{figure}

\subsection{Excluded-volume interactions for many particles in 1D}

Excluded-volume interactions, leading to emergent phenomena in $1$D for many interacting Lorenz-attractor-driven RTP-like particles (as shown in Figs.~$3$ and $4$ of the main text), were implemented as follows:

\begin{enumerate}
    \item On the unit interval, with periodic boundary conditions, we define the distance $d_{ij}$ between two particles $i$ and $j$ as the shortest distance between them:
    \begin{equation*}
    d_{ij}=\text{min}(|x_i-x_j|,1-|x_i-x_j|).
    \end{equation*}
    \item For each particle, we find the particles that are within a distance $d$, i.e.~$d_{ij}<d$. Here, the distance $d$ corresponds to the size of each particle.
    \item For each such pair of particles that are closer than the particle size, the motion ceases. However, their velocities continue to evolve on the strange attractor, so when a particle reverses its direction of motion and moves away from the other particle, the particle can resume its ballistic motion.
\end{enumerate}

An example $\mathtt{MATLAB}$ code, which implements these excluded-volume interactions for Lorenz-attractor-driven particles, can be found in the file $\mathtt{oneD{\_}many{\_}particles.m}$ of the Supplemental Material.

\subsection{Aligning interactions for two-dimensional flocks}

Aligning interactions, leading to the complex flocking dynamics in Fig.~\ref{Fig: 2D many} of the main text, were implemented as follows:

\begin{enumerate}
    \item For particles in the unit square with periodic boundary conditions, we define the shortest distance between the particles $i$ and $j$ as
    \begin{equation*}
    d_{ij}=\sqrt{(X_{ij})^2+(Y_{ij})^2},    
    \end{equation*}
where
    \begin{align*}
    X_{ij}&=\text{min}(|x_i-x_j|,1-|x_i-x_j|),\\ \nonumber
    Y_{ij}&=\text{min}(|y_i-y_j|,1-|y_i-y_j|).       
    \end{align*}
    
    \item For each particle, we define a neighborhood which is a circle of radius $\Delta$. If there is no other particle inside this neighborhood, then the particle can continue to move as an individual attractor-driven particle, driven by its strange attractor with its orientation $\theta_\textrm{own}(t)$.
    
    \item If there are other particles in the neighborhood of the $i$th particle, then we use a weight factor $W$ to determine the relative importance that the particle gives to the alignment of other particles, compared to its own directed motion driven by the strange attractor. Thus, the particle's orientation evolves according to
    \begin{align*}
      \theta_{\textrm{new}}(t)=W\,\theta_{\textrm{nb}}(t) + (1-W)\,\theta_\textrm{own}(t).  
    \end{align*}
    Here $\theta_{\textrm{nb}}$ is the average direction of the particles in the neighborhood, defined in the same way as the Vicsek flocking model~\cite{PhysRevLett.75.1226} (and as given in Sec.~\ref{Sec: AM 2D} of the main text). Note that we also include the direction of particle under consideration, when calculating the average.
\end{enumerate}

An example $\mathtt{MATLAB}$ code, which implements these aligning interactions for 2D Lorenz-attractor-driven particles, can be found in the file $\mathtt{twoD{\_}many{\_}particles{\_}flocking.m}$ of the Supplemental Material.

\section{One-dimensional active-matter-like attractor-driven particles}\label{app: 1D active particles}

\subsection{ABP-like attractor-driven particles}\label{ABPAOUP}

\begin{figure*}
\centering
\includegraphics[width=2\columnwidth]{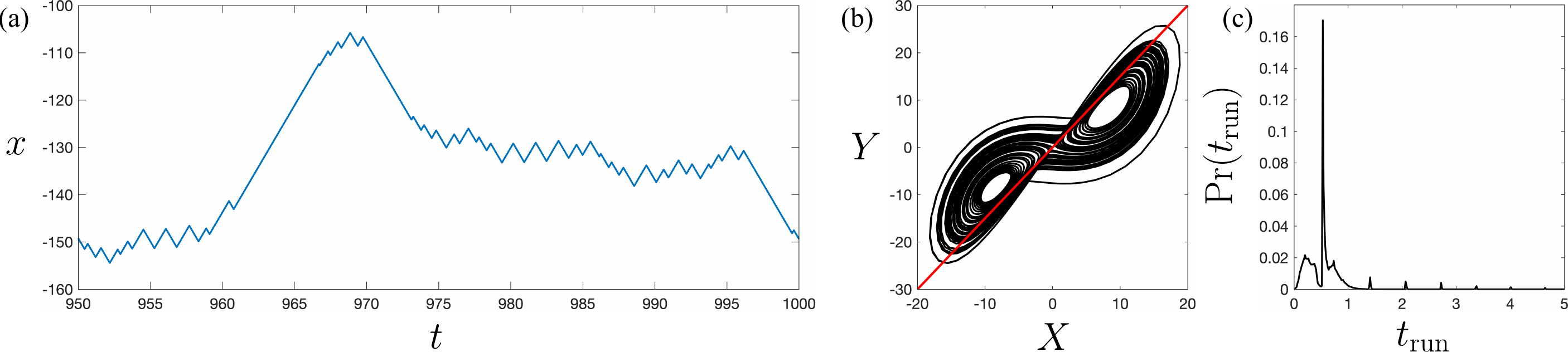}
\caption{One-dimensional Lorenz-strange-attractor-driven active particle with anomalous distribution of run durations. (a) Typical space-time trajectory. (b) The Lorenz attractor with parameters $\sigma=10, r=28, b=8/3$ drives the active particle moving in one-dimension with a constant speed $u=\sqrt{r-1}$. \textcolor{black}{The active particle reverses its} walking direction when the state-space trajectory on the attractor crosses the $Y=1.5 X$ plane. (c) Probability distribution of time spent in constant-speed ballistic motion between direction reversals.}
\label{Fig: 1D single ANM}
\end{figure*}

Using the formalism for generating one-dimensional RTP-like attractor-driven particles described in Sec.~\ref{Sec: AM 1D} of the main text, one can also generate an ABP-like particle in our attractor-driven framework. The equation of motion obeyed by a one-dimensional ABP is~\citep{D0SM00687D}
\begin{equation*}
    \dot{x}=u\,\cos(\theta(t)).
\end{equation*}
Here, $u$ is a constant speed, and $\theta(t)$ is an internal angular co-ordinate which in conventional ABP undergoes rotational diffusion. In our framework, we let $\theta(t)$ be determined from our internal state-space strange attractor. It is calculated at each time step using 
\begin{equation*}
\theta(t_{i+1})=\theta(t_i)+\Delta\theta(t_i). 
\end{equation*}
The internal state-space is a scaled Lorenz strange system given by
\begin{align*}
    \dot{X}&=F(\sigma(Y-X))\\ \nonumber
    \dot{Y}&=F(-Y+rX-XZ)\\ \nonumber
    \dot{Z}&=F(-bZ+XY),
\end{align*}
where $F$ is a non-zero real number. The trigger times $t_n$ are determined by the intersection of a phase-space trajectory along the scaled Lorenz attractor with the cutting plane $X=0$. The change in the internal angular co-ordinate, $\Delta\theta(t_i)$, is either (i) zero if the trajectory on the Lorenz strange attractor does not cross the cutting plane $X=0$ in the time interval between $t_{i-1}$ and $t_i$, or (ii) we select a value of $\Delta\theta(t_i)$ from a Gaussian distribution with a mean of $0$ and a standard deviation of $\sigma_\theta$ generated in a deterministic way (using the method described in Appendix~\ref{sec: arbit prob}) if the trajectory intersects the cutting plane during the time interval between $t_{i-1}$ and $t_i$. This gives an ABP-like motion in one dimension, whose internal angular co-ordinate undergoes rotational diffusion-like behavior with a Gaussian-like distribution. A typical trajectory for an ABP-like particle generated in this manner, together with its corresponding internal distribution of $\Delta\theta$, is shown in Fig.~\ref{Fig: ABP}. We note that although we have used our trigger times here as pseudo-random number generators to get a prescribed Gaussian distribution for the internal angular co-ordinate in order to mimic an ABP, our formalism is more general and allows us to generate an internal angle co-ordinate distribution inherent to the underlying internal state-space attractor by choosing an appropriate strange attractor and a cutting plane which may result in Gaussian-like or any other distribution arising naturally from the underlying attractor.

\subsection{1D attractor-driven RTP-like particle with anomalous distribution of run durations}

The Lorenz-attractor-driven RTP-like particle presented in Fig.~\ref{Fig: 1D RTP Lorenz} of the main text has an exponentially decaying envelope with discrete peaks in the probability distribution of run durations, which is analogous to the purely exponential distribution for a conventional RTP. However, we can also generate more diverse distributions of run durations for $1$D attractor-driven particles by changing the cutting plane and/or changing the underlying strange attractor. For example, as shown in Fig.~\ref{Fig: 1D single ANM}, keeping the same parameters of the Lorenz strange attractor as Fig.~\ref{Fig: 1D RTP Lorenz} of the main text but choosing the cutting plane to be $Y=1.5 X$ results in an anomalous probability distribution of run durations, where the particle has frequent short-run durations and occasional long-run durations.

\subsection{1D motile clusters of attractor-driven particles}\label{sec: motile clusters}

\begin{figure}
\centering
\includegraphics[width=\columnwidth]{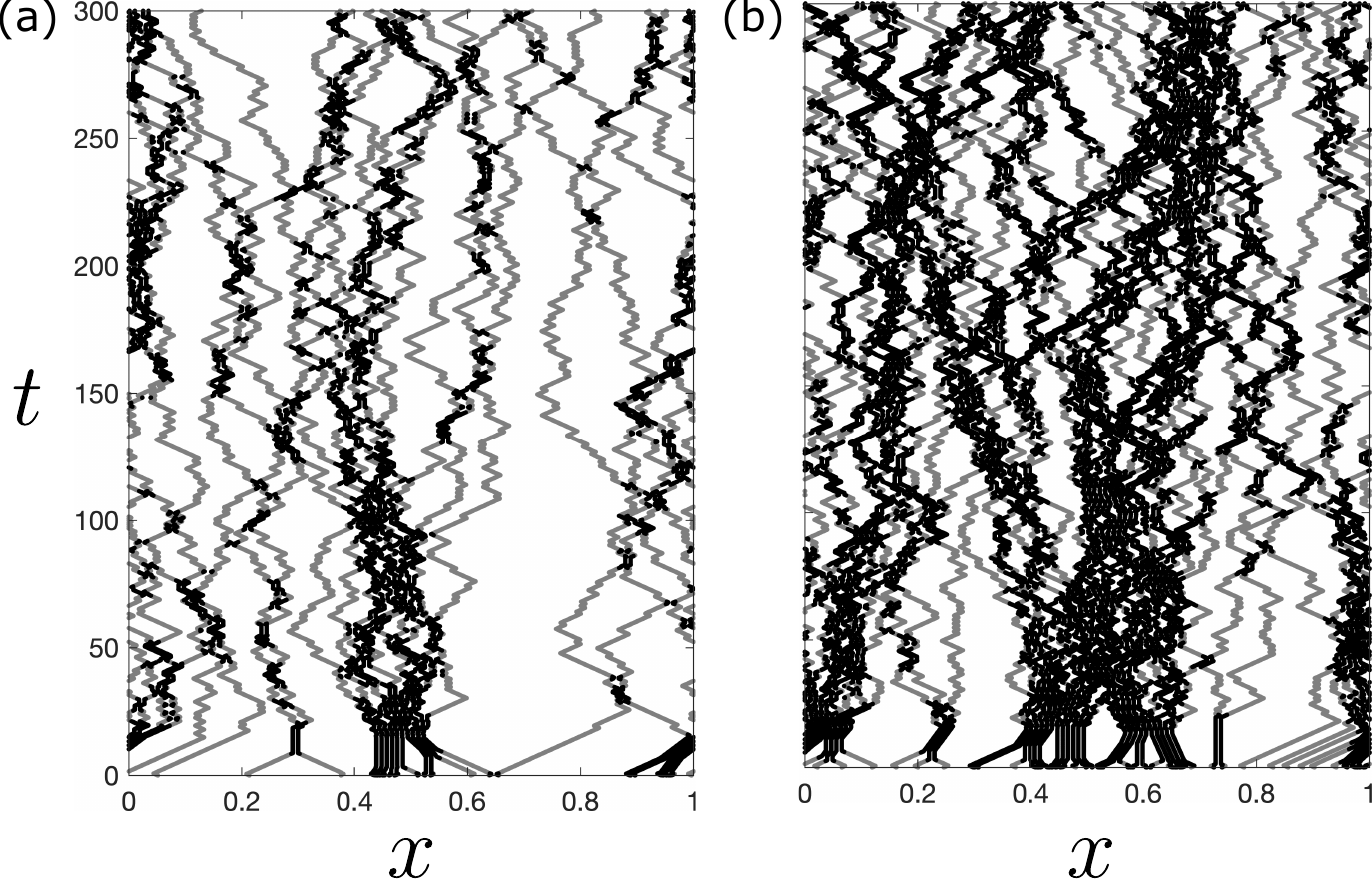}
\caption{1D motile clusters resulting from harmonic repulsive interactions between RTP-like Lorenz-attractor-driven particles. (a) Small motile clusters form which merge and disintegrate for a small number of particles $N=20$, while (b) larger more persistent clusters form for a larger number of particles $N=50$. Each particle has a speed of $u=0.01$, spring interaction constant $K=20$ and interaction distance $d_c=0.005$. The Lorenz attractor parameters are $\sigma=10$, $r=28$ and $b=8/3$. The gray trajectories denote isolated particles while the black trajectories denote clusters.}
\label{Fig: 1D motile cluster}
\end{figure}

In addition to jamming and cluster formation in one-dimension, as described in the main text, we can also generate one-dimensional motile clusters~\citep{D0SM00687D}. This can be done by including harmonic repulsive interactions between the one-dimensional RTP-like attractor-driven particles, instead of the excluded-volume interactions used for Figs.~\ref{Fig: 1D many} and \ref{Fig: 1D many 2} of the main text. We model the repulsive harmonic interactions between particles to activate when the distance between the particles falls below a certain threshold $2 d_{c}$, in which case the particles experience a repulsive force according to
\begin{equation*}
    F_{ij}=K\frac{x_i-x_j}{|x_i-x_j|} \left( 2 d_c - |x_i-x_j| \right).
\end{equation*}
We ensure that $|x_i-x_j|$ corresponds to the shortest distance between the particles in the periodic domain, and calculate the spring force accordingly. Figure~\ref{Fig: 1D motile cluster} shows a typical output of simulations where these interactions are modeled between the RTP-like particles. We observe that for a small number of particles, we obtain small-sized transient motile clusters, while for a larger density of particles, we see larger long-lived motile clusters.

\subsection{1D flocks of attractor-driven particles}

\begin{figure}
\centering
\includegraphics[width=\columnwidth]{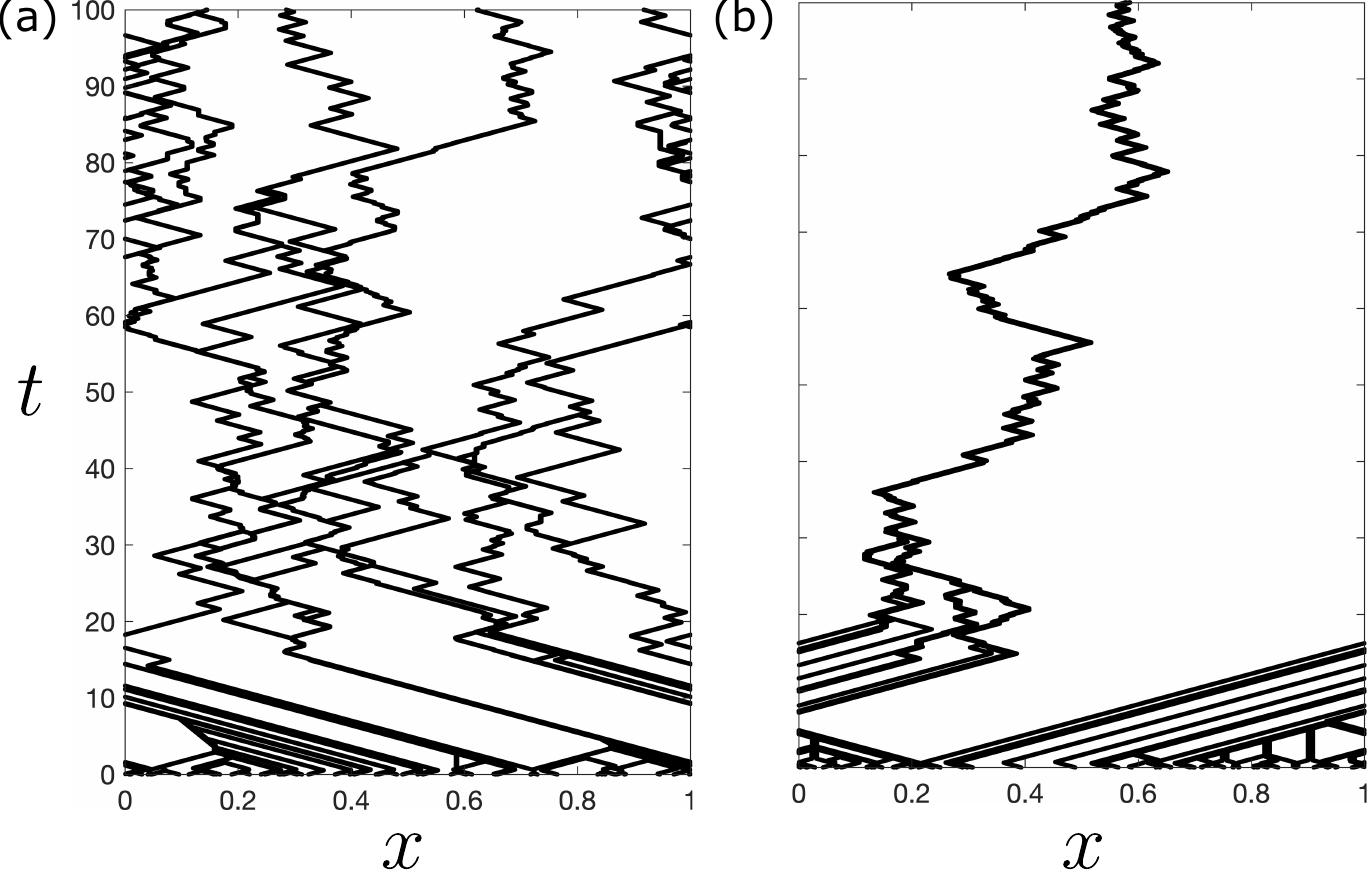}
\caption{1D flocking behavior from aligning interactions of many RTP-like Lorenz-attractor-driven particles. (a) Small flocks form which merge and disintegrate for a small flocking-interaction neighborhood of $\Delta=0.001$, while (b) a large single flock is formed for a large flocking-interaction neighborhood of $\Delta=0.008$. The larger single flock in (b) also intermittently changes direction. Each particle has a speed of $u=0.05$. The weight factor is $W=0.9$, and the number of particles is $50$. The Lorenz attractor parameters are $\sigma=10$, $r=28$ and $b=8/3$.}
\label{Fig: 1D flocking}
\end{figure}

We can generate one-dimensional flocking behavior~\citep{O_Loan_1999} by including aligning interactions for the one-dimensional RTP-like Lorenz-attractor-driven particles in a periodic domain. We implement the aligning interactions by allowing each such particle to detect other particles in a neighborhood of length $\Delta$ on either side of the particle. If there are no particles in the neighborhood, then the particle continues to evolve according to the RTP-like motion based on the Lorenz strange attractor. Conversely, if there are particles in the neighborhood then we sum the signs of the velocities of each of these particles (right as positive and left as negative) to find the direction of the majority of the particles in the neighborhood. For $i$th particle, this quantity is given by $\sum_{k\neq i} \text{sgn}(X_k)$. This quantity is then multiplied by a weight factor $W$ and added to the product of the particle's own velocity direction $\text{sgn}(X_i)$ and the corresponding weight factor $1-W$, giving us
\begin{equation*}
T_i=W \sum_{k\neq i} \text{sgn}(X_k) + (1-W)\,\text{sgn}(X_i). 
\end{equation*}
Then $\text{sgn}(T_i)$ determines the direction of motion of the $i$th particle. A typical example of the observed flocking behavior is illustrated in Fig.~\ref{Fig: 1D flocking}. Here we observe that for a smaller flocking-interaction neighborhood size of $\Delta=0.001$, we have a few small flocks while a larger flocking-interaction neighborhood of $\Delta=0.008$ results in a single larger flock that erratically changes direction.

\begin{figure}
\centering
\includegraphics[width=0.8\columnwidth]{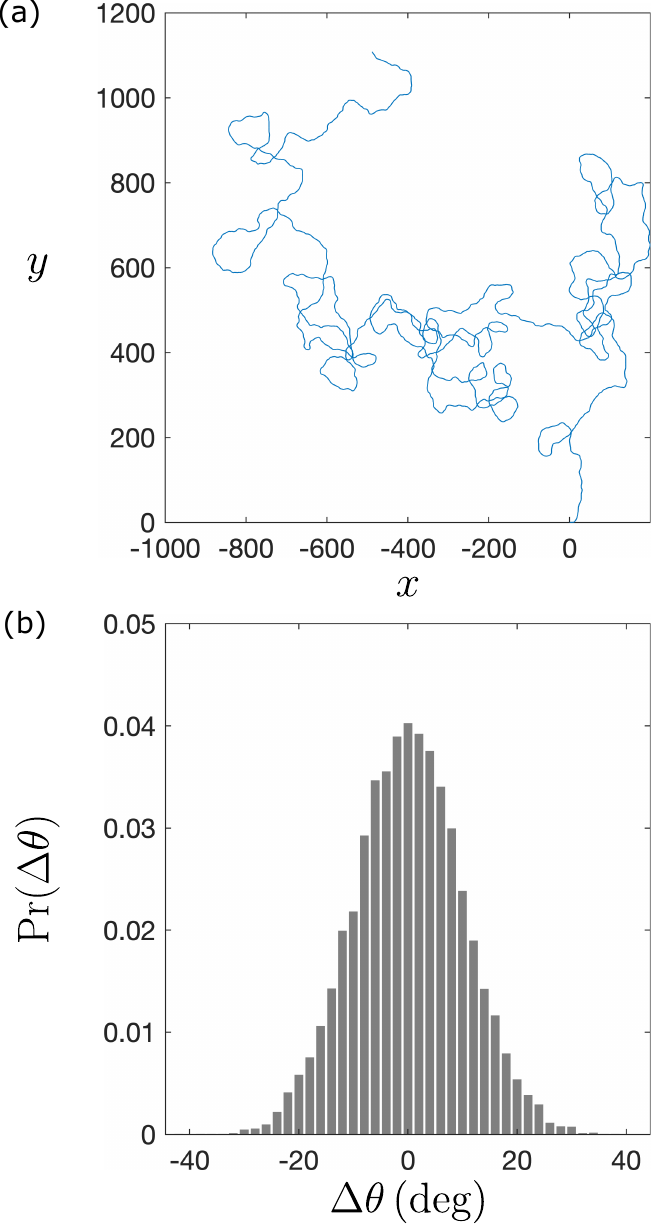}
\caption{Two-dimensional ABP-like particle generated from a scaled Lorenz system with parameters $\sigma=10, r=28, b=8/3$ and $F=10$. The trajectory of the particle is shown in (a), while (b) shows the probability distribution of the turning angle $\Delta\theta$. Here $u=\sqrt{r-1}$, the mean of the Gaussian distribution was chosen to be zero, and the standard deviation was chosen to be $\sigma_{\theta}=10^{\circ}$.}
\label{Fig: 2D ABP}
\end{figure}

\begin{figure}
\centering
\includegraphics[width=\columnwidth]{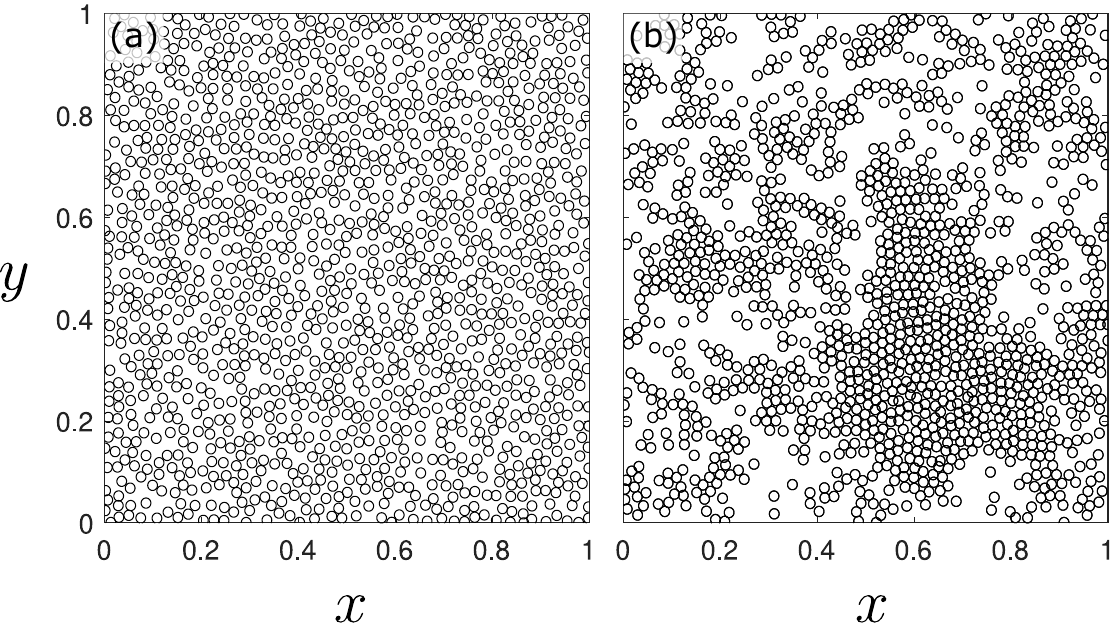}
\caption{(Multimedia view) Two-dimensional interactions of $1400$ ABP-like attractor-driven particles resulting in MIPS. (a) Initial state of the system where particles are started randomly, within a unit square with periodic boundary conditions. (b) Emergence of MIPS, where regions of high density and low density are formed due to the repulsive interactions between the ABP-like attractor-driven particles. The parameters for the ABP-like particles were the same as that described in Fig.~\ref{Fig: 2D ABP}. The spring constant for harmonic repulsive interactions was taken to be $K=50$ and the radius of each particle is $0.01$.}
\label{Fig: 2D ABPmany}
\end{figure}

\section{Generating arbitrary turning-angle distributions deterministically}\label{sec: arbit prob}

We can deterministically generate arbitrary probability distributions for turning angles, $\text{Pr}(\Delta\theta)$, using the trigger times $t_n$ as pseudo-random number generators (see Fig.~\ref{Fig: schemaic} of main text). Let
\begin{equation*}
\alpha=\pi (\phi -1), 
\end{equation*}
where 
\begin{equation*}
\phi=(1+\sqrt{5})/2 
\end{equation*}
is the golden ratio. Now take the trigger times $t_n$, divide by a numerical time-step size $\Delta t$ and round to the nearest integer, giving
\begin{equation*}
S_n=\text{round}\left(t_n/\Delta t\right). 
\end{equation*}
We then obtain an approximately uniform-distribution sampling of the turning angles $\Delta\theta_n$, using~\citep{Kingston:18} 
\begin{equation*}
\Delta\theta_n=S_n \alpha\,\text{mod}(2\pi).    
\end{equation*}
We can use this approximately-uniform distribution to generate an arbitrary distribution $\text{Pr}(\Delta\theta)$, using the inversion method described in Chap.~3 of the book by \citet{devroye:1986}. We start by discretizing the desired distribution, giving 
\begin{equation*}
P_i=\text{Pr}(\Delta\theta_i). 
\end{equation*}
Now, consider laying all these $P_i$ on a horizontal line in sequence as $P_1,P_2,...$, where the length of each line segment $P_i$ is its value. We can now generate an approximately uniform distribution to select a value on this line segment of length $\sum_i P_i$, according to 
\begin{equation*}
R_n=S_n \alpha\,\text{mod}\left(\sum_i P_i\right). 
\end{equation*}
We identify the line segment $P_j$ corresponding to length $R_n$ and sample the corresponding turning angle $\Delta\theta_j$. This allows us to deterministically sample numbers for an arbitrary distribution by using the trigger times $t_n$.

\section{Two-dimensional ABPs and motility-induced phase separation (MIPS)}\label{sec: 2D ABP}

Similar to the one-dimensional ABP-like motion described in Appendix~\ref{ABPAOUP}, we can also deterministically generate a two-dimensional ABP-like trajectory for a Lorenz-attractor-driven particle. We use the same scaled Lorenz system as in Appendix~\ref{ABPAOUP} and the same Gaussian distribution for the rotational diffusion. Implementing them with the following equation for two-dimensional particle velocity
\begin{align*}
\dot{x}=u \cos(\theta(t))\\ \nonumber
\dot{y}=u\sin(\theta(t)),
\end{align*}
we obtain a two-dimensional ABP-like trajectory as shown in Fig.~\ref{Fig: 2D ABP}. In conventional active matter, ABPs have been shown to undergo MIPS~\cite{doi:10.1146/annurev-conmatphys-031214-014710}. By exploring a large collection of two-dimensional ABP-like particles considered here, with added repulsive harmonic interactions as described in Appendix~\ref{sec: motile clusters}, we also obtain MIPS-like emergent behavior as shown in Fig.~\ref{Fig: 2D ABPmany}.

\bibliography{aipsamp}

\end{document}